\documentclass{article} 

\usepackage{preprint}
\usepackage{enumitem}
\usepackage{changepage} 
\usepackage{tabularx}
\usepackage{dcolumn}
\usepackage{makecell}
\usepackage{multirow}
\usepackage{hhline}
\usepackage{xspace}

\usepackage{siunitx} 
\usepackage[utf8]{inputenc}	
\usepackage[T1]{fontenc}	
\usepackage{xcolor}		
\usepackage[colorlinks = true,
            linkcolor = purple,
            urlcolor  = blue,
            citecolor = cyan,
            anchorcolor = black]{hyperref}	
\usepackage{nicefrac}		
\usepackage{microtype}		
\usepackage{float}			
\usepackage{multicol}		

\usepackage{newfloat}
\DeclareFloatingEnvironment[name={Supplementary Figure}]{suppfigure}
\usepackage{sidecap}
\sidecaptionvpos{figure}{c}

\usepackage{titlesec}
\titlespacing\section{0pt}{12pt plus 3pt minus 3pt}{1pt plus 1pt minus 1pt}
\titlespacing\subsection{0pt}{10pt plus 3pt minus 3pt}{1pt plus 1pt minus 1pt}
\titlespacing\subsubsection{0pt}{8pt plus 3pt minus 3pt}{1pt plus 1pt minus 1pt}

\ifpdf
  \pdfoutput=1\relax                   
  \usepackage{graphicx}                
  \DeclareGraphicsExtensions{.pdf,.png,.jpg,.jpeg} 
\else
  \usepackage{graphicx}                
  \DeclareGraphicsExtensions{.eps}     
\fi%

\PassOptionsToPackage{warn}{textcomp}  
\usepackage{textcomp}                  
\usepackage{mathptmx}                  
\usepackage{times}                     
\usepackage{cite}                      
\usepackage{tabu}                      
\usepackage{booktabs}                  

\usepackage{soul}
\usepackage{balance}
\usepackage{caption}
\usepackage{etoolbox}

\title{\textsc{SigmaCollab}: An Application-Driven Dataset \\for Physically Situated Collaboration}

\usepackage{authblk}

\author[1]{Dan Bohus}
\author[1]{Sean Andrist}
\author[1]{Ann Paradiso}
\author[1]{Nick Saw}
\author[2]{Tim Schoonbeek\thanks{Work conducted as part of an internship at Microsoft Research}}
\author[1]{Maia Stiber}

\affil[1]{Microsoft Research, Redmond, WA, USA}
\affil[2]{Eindhoven University of Technology, Eindhoven, Netherlands}
\affil[ ]{\small \texttt{\{dbohus;sandrist;annparadiso;nicksaw\}@microsoft.com; t.schoonbeek@tue.nl; maiastiber@microsoft.com}}

\newcommand{\insertfig}{\vspace{-5mm}\includegraphics[width=0.99\linewidth]{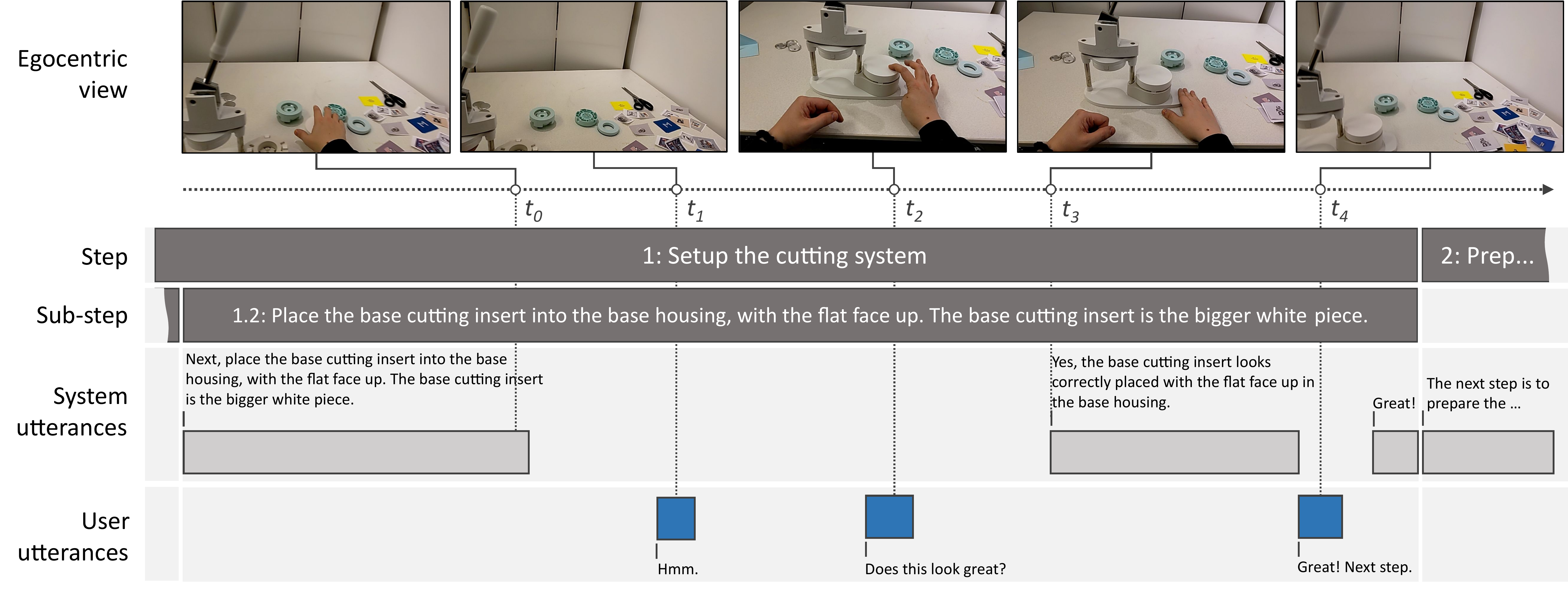}
\captionof{figure}{Schematic illustration of data streams from the \textsc{SigmaCollab} dataset in which the \textsc{Sigma} AI agent assists a participant to create a pin-back button using a button press machine. The transcript of the full interaction is shown in Appendix \ref{app:InteractionSample}.\vspace{7mm}}\label{fig:InteractionSample}}

\makeatletter
\apptocmd{\@maketitle}{\centering\setcounter{figure}{0}\insertfig}{}{}
\makeatother

\newcommand{\etc}{\textit{etc.}\xspace}
\newcommand{\eg}{\textit{e.g.}}
\newcommand{\ie}{\textit{i.e.}}

\begin{document}


\maketitle

\begin{abstract}

We introduce \textsc{SigmaCollab}, a dataset enabling research on physically situated human-AI collaboration. The dataset consists of a set of 85 sessions in which untrained participants were guided by a mixed-reality assistive AI agent in performing procedural tasks in the physical world. \textsc{SigmaCollab} includes a set of rich, multimodal data streams, such as the participant and system audio, egocentric camera views from the head-mounted device, depth maps, head, hand and gaze tracking information, as well as additional annotations performed post-hoc. While the dataset is relatively small in size ($\approx$ 14 hours), its application-driven and interactive nature brings to the fore novel research challenges for human-AI collaboration, and provides more realistic testing grounds for various AI models operating in this space. In future work, we plan to use the dataset to construct a set of benchmarks for physically situated collaboration in mixed-reality task assistive scenarios. \textsc{SigmaCollab} is available at \url{https://github.com/microsoft/SigmaCollab}.

\end{abstract}


\begin{multicols}{2} 

\section{Introduction}

Fluid human-machine interaction has been a decades-long pursuit in the research community. Building computational systems that can engage in interaction and collaborate with people \emph{in the physical world}, such as virtual assistive agents, interactive robots, or mixed-reality assistants brings forward multiple challenges at the intersection of several research fields: artificial intelligence, computer vision, natural language processing, and human-computer interaction. 

Physically situated systems need to make sense (in real time and to the extent needed by each specific application) of the surrounding physical environment and of the actions performed by humans in the environment. This often involves a set of competencies such as self-localization and mapping, object detection and tracking, depth estimation, face detection, body pose tracking, action recognition, and so forth. Many of these tasks have been extensively studied in the computer vision research community, with significant progress over the last two decades. The recent emergence of foundation models and large multimodal models has enabled a transition towards open-domain computer vision, and, by tying the models to language, has enabled reasoning about and answering more complex questions about the physical world. A lot of this progress has been made possible by the collection and curation of various datasets and data resources that have been used in the community to train and to comparatively evaluate the performance of various algorithms.

Situated collaboration raises, however, additional \emph{interaction-related} challenges. In addition to understanding the environment, objects, and actions, creating seamless situated collaborations requires fine-grained temporal coordination that is grounded in a deep understanding of human cognitive states such as, intentions, goals, beliefs, and various affective states. In these areas, progress has been slower. Today, most systems still use a "ping-pong" model of interaction where one user question is followed in a linear manner by one system answer, which limits naturalness and curtails opportunities for fluid, mixed-initiative collaboration. Part of the difficulty with interaction-related challenges lies in reifying, formalizing, and measuring the cognitive states and interactive phenomena involved. Moreover, the traditional approach of using collected, static datasets to build and evaluate models works only to a certain degree when it comes to interaction. Any trained model, once deployed, may alter user behavior, shifting the distribution of the underlying data. The study and evaluation of interactive phenomena are best done in the context of actual interactive systems. 

To foster and enable more research on interaction and collaboration challenges, we introduce a new resource: an \emph{application-driven, interactive dataset} called \textsc{SigmaCollab}. The dataset consists of 85 interactive sessions in which untrained participants interacted with a mixed-reality assistive AI application which guided them in performing certain tasks in the physical world, such as binding a notebook or installing the wheels on a skateboard. The dataset is egocentric and multimodal in nature; it includes audio, egocentric camera views, head, hands and gaze tracking, and it spans approximately 14 hours of interaction data.

\textsc{SigmaCollab}'s value-proposition lies in its \emph{application-driven} and \emph{interactive} nature. The data collection approach, in which participants interact with an AI application surfaces novel challenges that are not present in non-interactive datasets. As users engage with the system while pursuing meaningful and varied goals, the approach yields data with greater ecological validity. Finally, the open-source nature of the AI application used for collecting the data enables researchers to deploy models developed or evaluated based on this data in the context of the target application, and allows them to collect additional data to study their end-effects on metrics of overall task performance. We publish \textsc{SigmaCollab} to catalyze rigorous, application-grounded study of fluid human–AI collaboration and to close the gap between lab benchmarks and real-world performance.

\section{Related Work}
\label{sec:RelatedWork}

Publicly released academic datasets have historically played a transformative role in advancing computer vision research. The release of ImageNet~\cite{imagenet} was a turning point for large-scale visual classification, enabling the development of powerful deep convolutional architectures. Subsequently, COCO~\cite{coco} significantly pushed the field of object detection and segmentation, while Kinetics~\cite{kinetics} has become a cornerstone for video understanding and action recognition, and Cityscapes~\cite{cityscapes} for pixel-level scene understanding. Beyond these general-purpose datasets, application-driven benchmarks have similarly accelerated progress in specialized domains. For example, MNIST~\cite{mnist} catalyzed early breakthroughs in digit recognition and thereby modern deep learning, KITTI~\cite{kitty} has been instrumental in autonomous driving research, and LFW~\cite{lfw} in face recognition. These examples highlight that introducing a well-designed, meaningful dataset in a specific application area can catalyze rapid advances in methodology, benchmarks, and downstream impact that can shape the direction of entire research fields.

Spurred by an increased interest in embodied interaction, mixed-reality and robotics, multiple datasets and benchmarks have been developed in recent years for egocentric computer vision settings. For example, the OpenEQA dataset \cite{OpenEQA2023} provides embodied question-answer pairs that are anchored in videos (or 3D-scans) of real world environments. The EPIC-KITCHENS project \cite{Damen2018EPICKITCHENS, Damen2021PAMI, Damen2022RESCALING} captured daily activities in the kitchen from a first-person perspective, in support of challenges like action detection, action recognition, object segmentation, \etc The Ego4D dataset \cite{Ego4D2022CVPR} includes over 3000 hours spanning hundreds of different indoors and outdoors activities, and defines various benchmarks for memory, forecasting and social understanding challenges. Later, the Ego-Exo4D \cite{grauman2023egoexo4d} effort extends this by simultaneously capturing both ego and exo-centric views of skilled human activities and includes benchmarks for egocentric tracking of body and hand poses and for proficiency estimation. Other datasets like GTEA Gaze, GTEA Gaze+, EGTEA Gaze+ \cite{Fathi2012LearningTR, li2020eyebeholdergazeactions} focused on capturing eye gaze in conjunction with egocentric views, and leveraging that signal on various action-related challenges. A number of other egocentric datasets have also been constructed around structured, procedural task execution or performance (rather than spontaneous activities), such as Assembly101 \cite{sener2022assembly101}, CaptainCook4D \cite{peddi2024captaincook4ddatasetunderstandingerrors}, IndustReal \cite{schoonbeek2023industrealdatasetprocedurestep}, MECCANO \cite{EgoProceLECCV2022}, \etc These datasets support challenges like mistake detection, procedure step recognition, active object detection, action anticipation. 

While these egocentric datasets provide a rich empirical testbed for many computer vision challenges, they generally capture a single actor performing an activity in the world, and are therefore not supportive of interaction or collaboration-related challenges. In contrast, two other recent datasets, HoloAssist \cite{holoassist} and WTaG \cite{bao-etal-2023-foundation} were collected in an interactive setting: in both cases a human instructor guided a human worker who was wearing a HoloLens mixed-reality headset through performing a physical task. In addition to typical computer vision challenges, these data also surface a variety of interesting problems around interaction-related challenges, such as proactive interventions, grounding, reference generation and resolution, \etc While these interactive datasets are certainly valuable, both involve interactions between two humans (an executor and an instructor). In contrast, \textsc{SigmaCollab} is based on a stand-alone system that does not require human supervision. Consequently, its contents more accurately reflect challenges that arise in a realistic application.

\begin{figure*}[t!]
 \includegraphics[width=\textwidth]{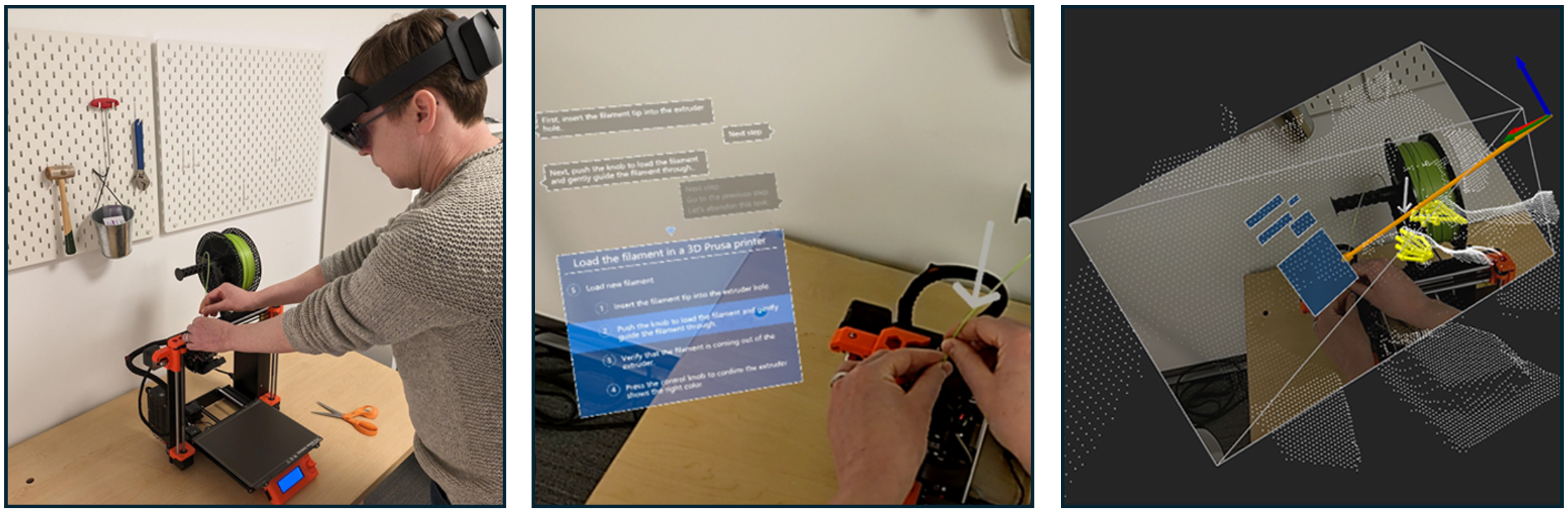}
 \caption{\textbf{Left}: user performing a procedural task with a mixed-reality headset running \textsc{Sigma}. \textbf{Middle}: first-person view showing \textsc{Sigma} guidance panel and task-specific holograms. \textbf{Right}: visualization of system's scene understanding showing the egocentric camera view, depth map, detected objects, gaze, hand and head pose in 3D space. \textcopyright \emph{2024 IEEE}}
 \label{fig:SigmaOverview}
\end{figure*}

\section{An Application-Driven Approach}

The datasets described above were collected using a diverse set of methods: scraping YouTube videos \eg \cite{Heilbron_2015_CVPR, miech2019howto100mlearningtextvideoembedding, zhou2017automaticlearningproceduresweb}, generating data using 3D simulators \eg \cite{OpenEQA2023, hazra2023egotvegocentrictaskverification}, capturing human activities with varying degrees of spontaneity or structure in lab \eg \cite{assembly101, Fathi2012LearningTR, li2020eyebeholdergazeactions, schoonbeek2023industrealdatasetprocedurestep, EgoProceLECCV2022} or in natural settings \eg \cite{Damen2018EPICKITCHENS, Damen2021PAMI, Damen2022RESCALING, Ego4D2022CVPR, grauman2023egoexo4d}, capturing human-human interactions \eg \cite{holoassist, bao-etal-2023-foundation}, \etc The annotations that often supplement the data and provide the basis for evaluation typically involve labor-intensive human effort, crowd-sourcing, and more recently the use of large language models. Each data collection and annotation approach colors the nature of the data, and problems that can be approached using it.

Our long term goal is to enable fluid human-computer interaction in physically situated settings. With this target in mind, we focus on the less-well addressed interaction-related challenges and we introduce an \emph{interactive, application-driven approach} to dataset development. Specifically, we collect data by having participants interact with an existing AI application. The participants are motivated by accomplishing an end goal, and the interactions, \ie, the data collected, are therefore situated in space and time within the context of a specific task. 

\sisetup{
  detect-all,
  input-decimal-markers = {.},
  table-format=3.1,
  separate-uncertainty = true,
  table-space-text-post = {\,kHz}, 
}

\begin{table*}[!b]
  \centering  
  \begin{tabularx}{\textwidth}{p{5cm}XSS}
  \toprule
   Stream & Representation & \multicolumn{1}{c}{Target} &  \multicolumn{1}{c}{Actual Avg.}\\
   &  & \multicolumn{1}{c}{Frame rate} & \multicolumn{1}{c}{Frame rate}\\
  \midrule
   Color Camera View& $896 \times 504$ pixels @ 24bpp, with camera pose and intrinsics & 15\,Hz& 14.91\,Hz\\
   Depth Camera View (long-throw) & $320 \times 288$ pixels @ 16bpp, with camera pose and intrinsics& 5\,Hz& 4.98\,Hz\\
   Left Front Grayscale Camera View & $640 \times 480$ pixels @ 8bpp, with camera pose and intrinsics & 15\,Hz & 13.64\,Hz\\
   Right Front Grayscale Camera View & $640 \times 480$ pixels @ 8bpp, with camera pose and intrinsics & 15\,Hz & 13.64\,Hz\\
   Head Pose + Eye Gaze & tuple of head pose matrix ($4 \times 4$) and eye gaze ray ($3 \times 1$ origin position vector and $3 \times 1$ direction vector) & 30\,Hz& 28.37\,Hz\\ 
   Hands Pose & pose matrices ($4 \times 4$) for each of the 26 joints in the left and right hand& 20\,Hz & 20.01\,Hz\\
   Audio & 1-channel, 32-bit floating-point PCM & 16\,kHz & 16.00\,kHz\\
  \bottomrule
  \end{tabularx}
  \caption{Configuration for \textsc{Sigma} streams: representation, target configured frame rates, as well as actual frame-rates during data collection.}
  \label{tab:SigmaSensorStreams}
\end{table*}

This approach has a number of advantages. First, it leads to collecting data that is ecologically more valid, in the sense that the challenges reflected in the data are better matched to the end goal of supporting fluid collaboration. For example, in an LLM-based approach to embodied question-answer generation, the constructed questions are most likely fully-formed sentences, such as \emph{"What is to the left of the microwave?"}, whereas the questions that naturally arise during an interaction may be fragments that contain a significantly larger number of referential expressions, such as \emph{"Like this?"} or \emph{"Is it this one?"}. At the same time, in contrast to data collected in human-human settings, the temporal patterns and content of utterances reflects the style people adopt when interacting with an AI agent, rather than with another human. 

Second, the approach can bring to the fore new questions and challenges. For example, one observation in the \textsc{SigmaCollab} dataset collected with this approach is that participants often talk to themselves throughout the task execution. This in turn raises the challenge of detecting self-talk, \ie, understanding which utterances should be responded to and which not. This is a interaction-related competency that an agent should have, but that to our knowledge has not been addressed in the community.

Finally, because the framework relies on an open-source application, models developed or evaluated on this data can be integrated back into the original application, and tested in new live interactions to understand the full, end-to-end effects on task-level performance and user satisfaction metrics. This in turn can surface new questions and challenges, within successive cycles of refinement.

In this work we use this interactive, application-driven approach to construct a dataset for physically situated collaboration. We use \textsc{Sigma} \cite{sigmaarxiv, sigmaieeevr} as the target application  --- an open-source \cite{sigma_github}, mixed-reality task assistive research testbed system that guides the participants through the execution of various procedural tasks in the physical world. In the next section we describe \textsc{Sigma} in more detail, including the system architecture and the data capture configuration. Next, in Section \ref{sec:Dataset}, we describe the data collection experiment and present basic dataset statistics. In Section \ref{sec:Annotations} we discuss additional annotations and post-hoc processing performed on the data. Finally, in Section \ref{sec:Conclusion} we present conclusions and outline future work.

\section{\textsc{Sigma}}
\label{sec:Sigma}

We begin with a brief description of \textsc{Sigma}'s functionality and system architecture (an in-depth description and additional technical details are available in \cite{sigmaarxiv}). Next, we outline a number of adaptations made to the system in subsection \ref{subsec:SigmaUpdates} and discuss the specific configuration used during the data collection in subsection \ref{subsec:SigmaConfiguration}. 

\subsection{Overview}

\textsc{Sigma} is an open-source \cite{sigma_github} system that was developed to serve as a testbed for research in mixed-reality task assistance \cite{sigmaarxiv, sigmaieeevr}. The system leverages a HoloLens 2 headset and can guide users step-by-step through performing procedural tasks in the physical world (see Figure \ref{fig:SigmaOverview}).

Task recipes are organized linearly, into a series of steps with sub-steps, and can be generated automatically via an LLM or can be manually authored in a simple json format. Throughout the interactions, the system displays a virtual panel containing the task instructions, as shown in Figure \ref{fig:SigmaOverview}, and reads aloud the instructions to the user one sub-step at a time. A history of the recent dialog is visible to the user in a floating virtual panel above the task recipe (see also Figure \ref{fig:SigmaOverview}). In addition, \textsc{Sigma} may also render step-specific holograms, such as the downward pointing arrow shown in middle image from Figure \ref{fig:SigmaOverview}. Once the user completes a sub-step, they can navigate to the next sub-step with a simple voice command like \textit{``Next step''}. Users can also ask questions throughout the interaction, and other navigational commands allow going directly to a particular step and sub-step number or going backwards, to a previous step.

Figure \ref{fig:InteractionSample} illustrates a snippet from a larger interaction in which a user is using a button press machine to create a pin-back button (the entire exchange is presented in Appendix \ref{app:InteractionSample}). Once the system provides a sub-step instruction to the user, after each user utterance \textsc{Sigma} first uses keyword spotting over the speech recognition results to identify simple user requests like \textit{``next``} or \textit{``previous step``}. If the keyword spotting process fails, the system relies on an LLM to decide how to respond to the user. Specifically, a \emph{response-generating prompt} that leverages the dialog context, as well as visual information (the egocentric image at the beginning of the user utterance) is used to instruct an LLM to classify the last user utterance into one of four classes: \texttt{question} (the user is asking a question), \texttt{no-question} (self-talk, \ie, the user is talking or making a comment to themselves), \texttt{next-step} (the user is making a statement indicating that they are ready to move on to the next step), or \texttt{step-navigation} (the user is issuing a more complex navigation command); additionally the prompt also instructs the LLM to provide an answer if the utterance is classified as a question. An example of the an instantiated prompt and corresponding response is shown in Appendix \ref{app:PromptSample}.

Architecturally, \textsc{Sigma} is implemented as a distributed system: a lightweight client application runs on the HoloLens 2 device configured in Research Mode \cite{hololens_researchmode}, and communicates live with a server application that runs on a desktop computer. The client application captures data from the various HoloLens sensors (\eg, color, depth in long-throw mode, grayscale cameras, microphone) and streams them live to the server, which runs the bulk of the computation, including speech recognition, interaction management, speech synthesis, \etc The server application streams back to the client in real time instructions for actuating the user interface, and rendering holograms and audio. The on-device sensor streams, together with additional streams of data generated by the server, are persisted on the server side. For the interested reader, additional technical details of the system's architecture and implementation are available in \cite{sigmaarxiv}.

\begin{figure*}[p!]
    \begin{center}
    \includegraphics[width=\textwidth]{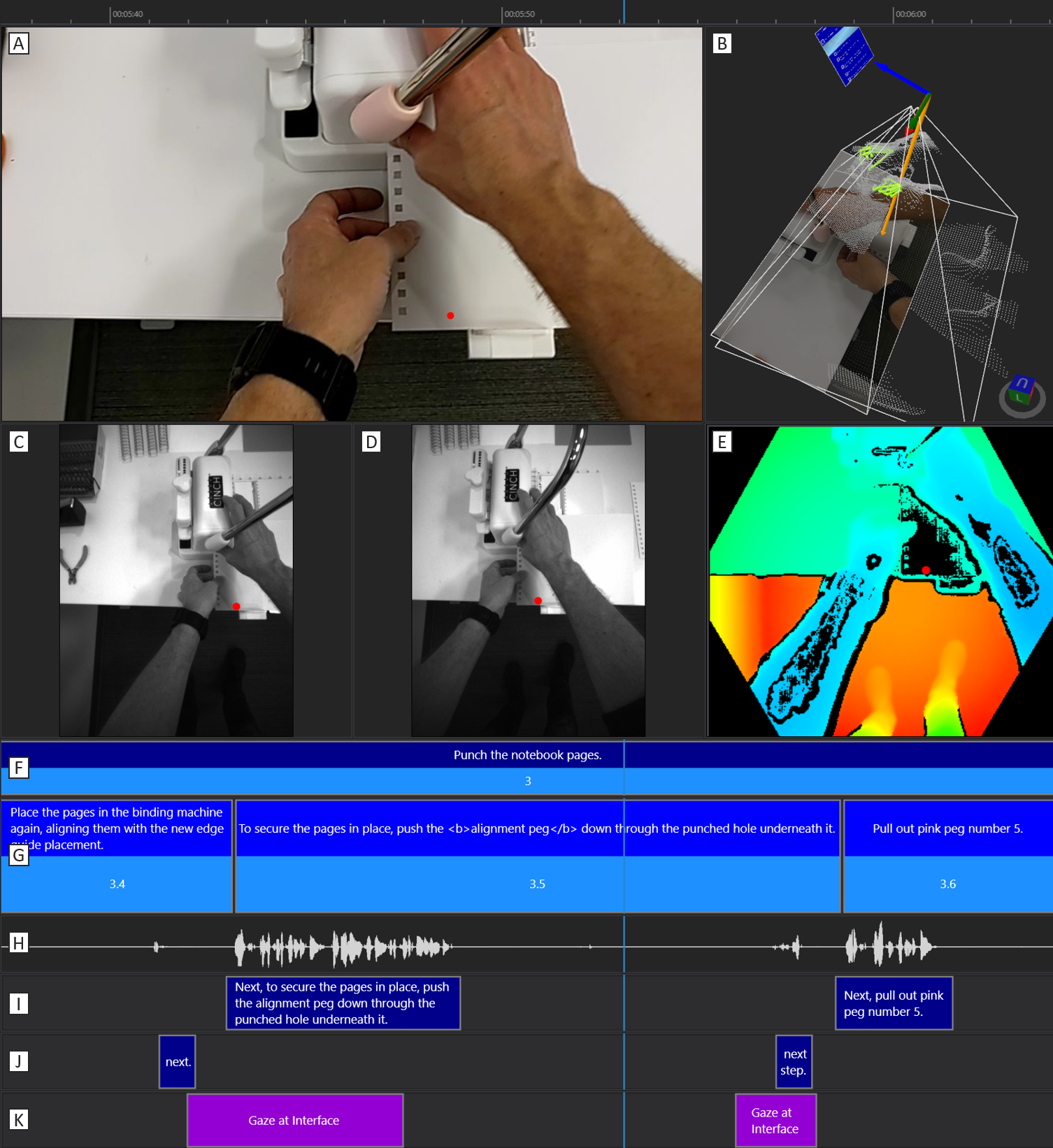}
    \caption{Illustration of the various data streams available in the \textsc{SigmaCollab} dataset. Top of the image has the time axis and the blue vertical line represents the current time position for the image streams. (A) Color camera image; (B) 3D color and depth camera poses, head pose, gaze direction, hands, and mixed reality UI positioning; (C) left grayscale camera image; (D) right grayscale camera image; (E) depth image; (F) temporal step execution boundaries; (G) temporal sub-step execution boundaries; (H) audio; (I) system utterances; (J) participant utterances; (K) temporal intervals in which the participant is gazing at the mixed reality user interface panel. The red dots in (A), (C), (D), and (E) show the corresponding projected gaze points.}
    \label{fig:PsiStudioCapture}
    \end{center}
\end{figure*}

\subsection{System Updates and Changes}
\label{subsec:SigmaUpdates}

Prior to data collection, we performed a number of pilot experiments with \textsc{Sigma} and made a several changes and adjustments to the system based on lessons learned. Some, but not all, of these experiments are described in \cite{bohus2024isitecologicallyvalid}.

The original version of \textsc{Sigma}, described in \cite{sigmaarxiv}, used a text-only response-generating prompt. We updated the system to leverage multimodal models (such as the GPT-4o family of models) and use both text and images when responding to the user. We designed and tuned a more complex response-generating prompt that includes information about the task recipe, the step and sub-step the user is currently on, the history of the dialog in the current sub-step, as well as an image collected at the beginning of the user utterance from the right-side egocentric grayscale camera on the device. We chose one of the egocentric grayscale cameras for this because its field of view better captures the user's hands than the color camera in the HoloLens2 device, which is pointed more upward and often misses important portions of the workspace. More details about the design of this prompt, together with an example are shown in Appendix~\ref{app:PromptSample}.

To improve response latencies, we transitioned from the Azure Speech~\cite{azure_speech} service used by default in \textsc{Sigma} and implemented a solution that combines the Silero voice activity detector \cite{silero_github} with the Whisper speech recognition model \cite{whisper}. For Whisper, we selected the smallest and fastest variant, \texttt{tiny.en}, to optimize latency. We also implemented a continuation detector: it detects when a user begins speaking shortly after pausing, and merges the speech fragments into a single utterance for processing. In addition, we implemented support for canceling LLM requests that are in flight if the user starts speaking before the results from the LLM arrive. In this case, LLM query triggered by the first user utterance or fragment is canceled and a new query is issued. 

\begin{table*}[t]
  \centering
  \begin{tabular}{llccc}
  \toprule
    Task & Description & \# Steps & \# Sub-steps & Duration \\
    & & & & (mm:ss) \\
    \midrule
     Nespresso & Make coffee using a Nespresso Pixie machine  & 4 & 10 & 03:33 \\
     Hard-drive & Replace a hard-drive in a PC  & 4 & 12 & 03:50 \\
     Skateboard & Install the wheels on a skateboard & 5 & 23 & 10:26 \\
     Button & Make a button using a button press machine  & 6 & 27 & 08:40 \\
     Notebook & Make a notebook using a binding machine  & 5 & 34 & 10:07 \\
     Margarita & Make a Margarita mocktail & 3 & 14 & 06:28 \\
     Mojito & Make a Mojito mocktail  & 4 & 17 & 07:39 \\
     Whiskey & Make a Whiskey Sour mocktail  & 5 & 18 & 07:44 \\
    \bottomrule
  \end{tabular}
  \caption{Tasks, together with number of steps and sub-steps and duration in the expert demonstration.}
  \label{tab:Tasks}
\end{table*}

Finally, we made several other performance-focused changes to optimize the streaming of sensor data from client to desktop. Other minor changes included improvements to support for step navigation, and minor user interface updates.

\subsection{Data Capture Configuration}
\label{subsec:SigmaConfiguration}

We configured the \textsc{Sigma} client app to collect image views from the RGB, depth and the two front-facing grayscale cameras, at the resolutions and target frame-rates described in Table \ref{tab:SigmaSensorStreams}. For each camera, we also captured the camera intrinsics parameters, as well as the live camera position in 3D-space (relative to the same world anchor created in the lab used for data collection). Additionally, we captured audio from the device microphone array, as well as head, hand and gaze tracking information. The sensor data streams described in Table \ref{tab:SigmaSensorStreams} provide a rich multimodal basis for constructing future challenges for situated collaboration. Figure \ref{fig:PsiStudioCapture} shows a sample visualization of the various streams available in the data.

Three different Azure GPT model deployments were used in conjunction with the response-generating prompt throughout the data collection study. In the first part of the study we relied on a GPT-4o model using a global data zone region\footnote{geographical region for the datacenter providing the service}. After observing relatively large response latencies, we switched to a GPT-4o-mini deployment in a US data zone region. Finally, for the last third of the dataset, we switched back to a GPT-4o deployment using a US data zone region. The resulting three sections of the dataset are described in Appendix \ref{app:DatasetSections}.

\section{\textsc{SigmaCollab} Dataset}
\label{sec:Dataset}

To assemble \textsc{SigmaCollab}, we conducted a data collection experiment in which participants interacted with \textsc{Sigma} to perform a set of procedural tasks. The experiment was approved by the Microsoft Research Institutional Review Board. Below, we describe the battery of tasks used, the data collection protocol and relevant statistics from the resulting dataset.

\subsection{Tasks}
\label{subsec:SigmaTasks}

We defined eight procedural tasks (see Table \ref{tab:Tasks}) to be used in the data collection study. The tasks span multiple domains. They include a simple task with a short number of sub-steps: making coffee using a Nespresso machine; two hardware tasks: installing wheels on a skateboard and changing a hard-drive in a computer; two craft tasks: making a pin-back button and making a notebook; and three mocktail preparation tasks: making a Margarita, making a Mojito, and making a Whiskey Sour. 

We designed the task recipes in an iterative process, and brought further refinements based on observations made in pilot studies. We did not include any step-specific holograms in the task definitions, hence the only holograms visible to the user throughout the task execution are the task panel and dialog history. The final set of recipes, together with some illustrative image captures for each task, are shown in Appendix \ref{app:TaskRecipes}.

The complexity of the tasks varies, as reflected by the number of steps and sub-steps in each recipe (see Table \ref{tab:Tasks}): the simplest task has a total of 10 sub-steps, and the longest has 34. When executed by an expert that simply performs each of the sub-steps correctly, at a regular pace, and without asking any questions, the durations vary from 3.5 minutes to over 10 minutes.

The tasks span a diverse range of objects, ingredients, materials, and tools. These items vary along several dimensions: 

\begin{itemize}
    \item Size: from very small (\eg, screws, buttons) to medium-sized (\eg, skateboard, Nespresso machine);
    \item Familiarity: some are common, everyday items (\eg, screwdriver, lemon, coffee cup), while others are specialized tools unfamiliar to most people (\eg, skate tool, cinching machine, Hawthorne strainer);
    \item Quantity: some are single items (\eg, Nespresso machine, Boston shaker), others appear in multiples (\eg, ice cube, Nespresso capsule, mint leaf), and some are treated as sets (\eg, ice cubes, mint leaves);
    \item Material properties: the tasks involve both non-deformable and deformable objects, including various liquids.
\end{itemize}

Furthermore, not all objects present in the workspace are relevant to the task at hand. Overall, these dimensions of variability are similar to what one might encounter in real-world interactive scenarios, and increase the computer vision challenges present in this dataset.

The steps in the selected procedural tasks also exhibit a range of characteristics involving various kinds of physical actions to manipulate the relevant task objects. There are several different types of steps present across all eight tasks (the full mapping of all task steps to their specific category can be found in Appendix~\ref{app:TaskRecipes}): 

\begin{figure*}[t!]
    \includegraphics[width=\textwidth]{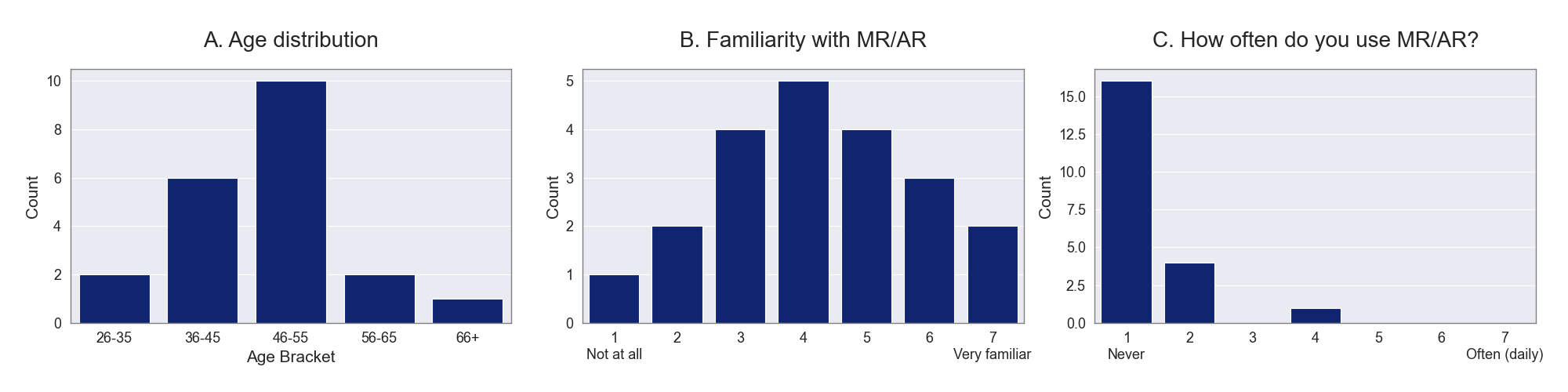}
    \caption{Data collection study participant demographics.}
    \label{fig:Demographics}
\end{figure*}

\begin{itemize}
\item Discrete Event: characterized by a crisp, instantaneous event that marks the successful completion of the step, \eg, ``Plug the wide connector into the hard-drive.''

\item Move: an object is moved, removed, or placed in some position, possibly involving fuzzy micro-adjustments after the placement, \eg, ``Take a front and back cover and place them in the binding machine, aligning with the edge guide.'' This step might require small adjustments to accomplish the final alignment after placing.

\item Add Amount: an incremental or continuous pouring, squeezing, or addition of a substance into a container, usually a liquid, \eg, ``Pour in 0.5 ounces of simple syrup.''

\item Check: a conditional adjustment that may or may not involve action, \eg, ``Adjust the kingpin nut if necessary to achieve the desired tightness or looseness for turning.''

\item Invisible State: a step where the final outcome is not directly observable, and is only inferable by the users’s completion of their action, \eg, ``Screw the case screw back using the screwdriver.'' It is not directly observable from visible object relationships when the screw is fully screwed into the case, but only indirectly inferable by virtue of the user ending their screwing action.

\end{itemize}

\subsection{Data collection protocol}

The study was performed in a laboratory setting, where we set up workstations for each of the eight tasks described in the previous section. There was one workstation dedicated to each of the five non-mocktail tasks, and a sixth workstation for all three mocktail tasks; this latter workstation was equipped with a super-set of the necessary tools and ingredients for all mocktail recipes. The images at the beginning of each task recipe in Appendix \ref{app:TaskRecipes} illustrate the various workstation setups.

We recruited participants for the study from among the co-workers at our organization, via broadly reaching emails and word-of-mouth. In total 21 participants engaged in the data collection study and provided permission for public data release. Each data collection session lasted up to ninety minutes, and participants were compensated 75 USD for their time and participation. 

We used the following protocol with each participant in the data collection study. Upon arriving at the lab, the participant was greeted by the researcher running the study, and a brief introduction to the overall goals of the study was presented. After signing an informed consent and data release form, the participant watched a 5-minute slide deck to familiarize themselves with the HoloLens device and how the \textsc{Sigma} system works; the deck included a short video with a first-person view of \textsc{Sigma} in action (on a different task than the ones used throughout the study). Next, the headset was placed on the participant, and an eye calibration process was performed. The participant was then given a final set of instructions: they were asked to do their best to complete each task, but were also informed that they could abandon a task at any point they felt unable to make progress or started to experience physical or mental discomfort; they were instructed not to interact with the researcher conducting the study; finally, they received a set of safety guidelines.

Following this briefing, each participant engaged in performing up to 6 tasks chosen from a broader set of predefined procedural tasks. The tasks were randomly chosen for each participant, following a stratified approach: a random order was created between the 6 workstations in the lab corresponding to the 5 non-mocktail tasks and the general mocktail category, and the choice of which mocktail to perform was also randomly chosen. 

Throughout the task execution, the researcher conducting the study was separated visually (behind a panel) from the participant, but was able to monitor the interaction remotely, through live visualization of the data streams from the device enabled by the PsiStudio tool \cite{psi_arxiv, psi_rapid}. In case of application crashes (a few happened occasionally), the experimenter intervened to inform the participant about the application failure and instruct them to move on to the next task. In addition, the experimenter intervened and ended the task in situations where it was clear that the participant was completely blocked from making further progress after committing an unrecoverable mistake. For example, the experimenter intervened if the participant accidentally triggered the Nespresso machine's cleaning cycle, broke their pin-back button after pressing it incorrectly, started installing their skateboard components to the completely wrong part of the deck, and so on.

After each task, the experimenter decided if enough time was left to perform the next task. If so, the participant was instructed to proceed with the next task. If not, the experiment was brought to a conclusion and the participant was debriefed and asked to fill in a demographics questionnaire (shown in Appendix \ref{app:DemographicsQuestionnaire}).

\subsection{Demographic information}

Of the 21 participants, 12 were male and 9 female. Additional demographic information from the post-experiment questionnaire is shown in Figure \ref{fig:Demographics}. Most participants were in the 46-55 age bracket (Figure \ref{fig:Demographics}.A). While their level of familiarity with AR/VR technologies varied, most participants had already encountered these technologies (Figure \ref{fig:Demographics}.B), even if they were not using them often (Figure \ref{fig:Demographics}.C)

\subsection{Dataset statistics}

We denote by \emph{task execution session} or in short \emph{session} a participant's attempt to execute one of the tasks, regardless of whether the task was performed correctly. The 21 participants attempted a total of 95 task execution sessions. Of these, 10 sessions were excluded due to application crashes and performance issues. The \textsc{SigmaCollab} dataset therefore consists of the remaining 85 task execution sessions. Together, these sessions span a total duration of 13 hours, 45 minutes and 11 seconds. Throughout the 85 sessions, the participants attempted 1583 sub-step executions and produced 3296 utterances.

\begin{figure*}[t]
    \includegraphics[width=\textwidth]{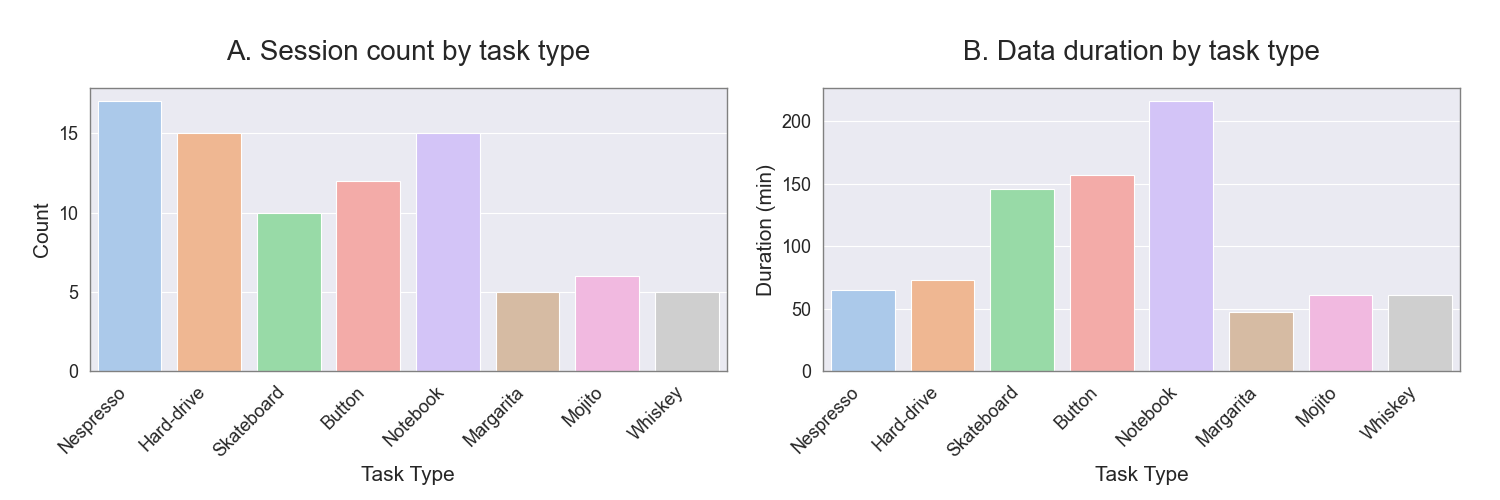}
    \includegraphics[width=\textwidth]{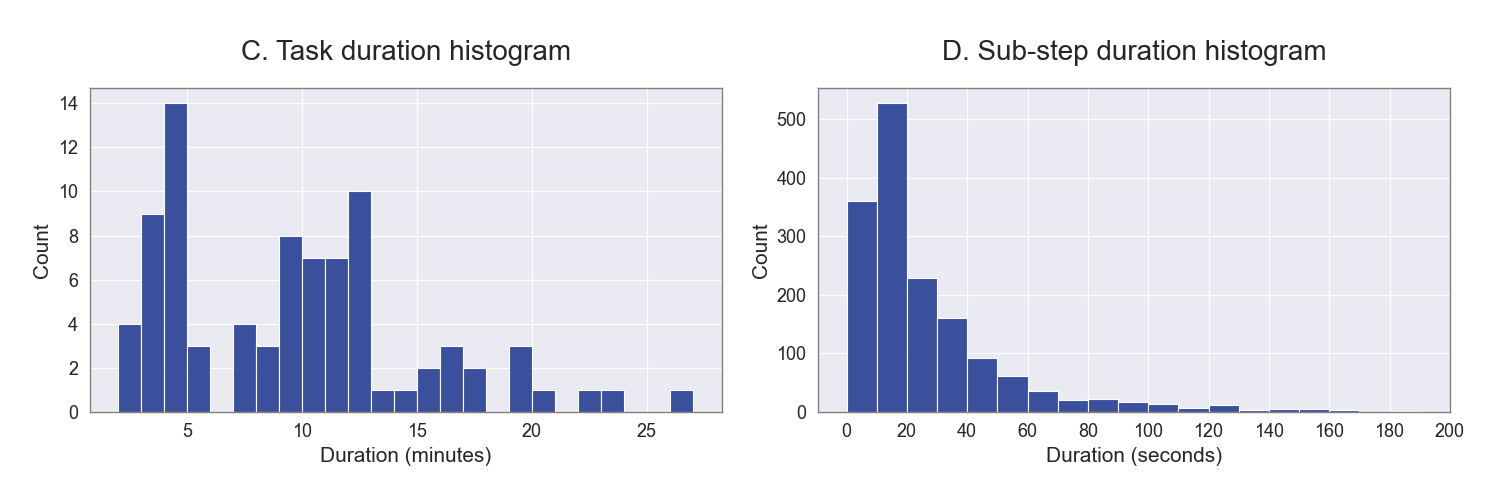}
    \caption{Dataset statistics.}
    \label{fig:DatasetStatistics}
\end{figure*}

Several additional statistics are presented in Table \ref{tab:DatasetStatistics}, and illustrated in Figure \ref{fig:DatasetStatistics}. The attempted task executions varied significantly in duration: the shortest attempted task execution in the dataset was 2 minutes and 1 second, and the longest 26 minutes and 11 seconds; the distribution of task durations is shown in Figure \ref{fig:DatasetStatistics}.C. We eliminated from the analysis sub-step executions in which the participant decided to abandon the task, or where the experimenter interrupted to stop the experiment, as described in the previous sub-section. The remaining 1570 attempted sub-step executions vary in duration, ranging from 2.16 seconds to 7 minutes and 53 seconds, with an average of 28.6 seconds; the distribution of sub-step durations is shown in Figure \ref{fig:DatasetStatistics}.D.

\subsection{Expert Demonstrations}

\begin{table*}
  \centering
  \begin{tabular}{lccccc}
  \toprule
         & Task Execution & Total Duration & Sub-step & User Utterance & \multicolumn{1}{c}{Avg. Duration} \\
    Task & Sessions (\#) & (hh:mm:ss) & Executions (\#) & (\#) & \multicolumn{1}{c}{(mm:ss)} \\
    \midrule
     Nespresso & 17 & 01:05:16 & 148 & 295 & 03:50 \\
     Hard-drive& 15 & 01:12:41 & 178 & 288 & 04:51 \\
     Skateboard & 10 & 02:25:20 & 218 & 540 & 14:32 \\
     Button & 12 & 02:36:33 & 296 & 619 & 13:03 \\
     Notebook & 15 & 03:35:37 & 480 & 943 & 14:22 \\
     Mocktail & 16 & 02:49:45 & 263 & 611 & 10:36 \\
     \hspace{0.5cm}Margarita & 5 & 00:47:27 & 72 & 161 & 09:29 \\
     \hspace{0.5cm}{Mojito} & 6 & 01:01:05 & 99 & 222 & 10:11 \\
     \hspace{0.5cm}{Whiskey} & 5 & 01:01:13 & 92 & 228 & 12:15 \\
    \midrule
    Overall (Dataset) & 85 & 13:45:11 & 1583 & 3296 & 09:46 \\
    \bottomrule
  \end{tabular}
  \caption{Dataset statistics, overall and by task type.}
  \label{tab:DatasetStatistics}
\end{table*}

In addition to the 85 task execution sessions collected from the 21 participants, we also collected 8 \emph{expert demonstration} sessions, one for each type of task, performed by a user who was very familiar with the tasks (one of the authors). During these expert demonstration sessions the user performed each step correctly, at a regular pace, and without asking any questions. As expected, these sessions are all successful and their durations are generally shorter than the corresponding task execution sessions from the participant data (as seen in Table \ref{tab:Tasks} for expert demonstrations and Table~\ref{tab:DatasetStatistics} for participant data). The expert demonstration sessions are released together with the main dataset, and can be used in one-shot learning scenarios, or can provide useful prior information for a variety of downstream tasks and challenges.

\subsection{Dataset Splits}
The dataset is primarily intended to serve as a testbed for evaluating the generalizability and effectiveness of various models in the context of an interactive application, which is potentially out-of-distribution from what the models have been trained on. As such, we do not define specific train/val/test splits. That being said, specific train/test/val splits can be created based on the problem space investigated. When generating such splits, care should be exerted around stratification: the dataset can be stratified based on various data attributes, including but not limited to dataset sections (see Appendix \ref{app:DatasetSections}), participant ids, task types, step types, task success, and combinations thereof.

\section{Annotations and Post-processing}
\label{sec:Annotations}

Next, we describe the set of manual annotations and additional post-processing that we performed on the collected data. 

\subsection{Manual Transcriptions}
\label{subsec:ManualTranscriptions}

Throughout the data collection, \textsc{Sigma} used the Silero \cite{silero_github} voice activity detector in conjunction with the \texttt{tiny.en} version of the Whisper \cite{whisper} speech recognizer. Since both these models sometimes made errors, we manually segmented and transcribed the utterances spoken by the user, and provide the reference transcripts as part of the dataset release. 

To assess the runtime performance of the Silero voice activity detector, we computed the detection error rate, \ie~the duration of falsely detected speech segments plus the duration of missed speech segments, divided by the total duration of the reference (manually transcribed) speech segments. The results indicate an average session-level detection error rate of 12.4\%. To assess the runtime performance of the Whisper speech recognizer, we computed session-level word-error-rate by concatenating all the speech recognition results and all the manual utterance transcriptions from a session and computing word-error-rate on the resulting strings. The average session-level word-error-rate was 20.2\%. These numbers confirm that the runtime speech recognition results contain significant errors, and the manual transcripts can provide an important resource for various downstream challenges and analyses.

\subsection{Word-Level Timestamps}
\label{subsec:WordLevelTimestamps}

Whisper can be configured to provide word-level timestamps. However, our initial experimentation revealed that these timestamps are often inaccurate. We therefore computed word-level timings in post-processing, by using a force-alignment algorithm \cite{forced_alignment_algo, forced_alignment_pytorch}. The word-level timings were computed for:

\begin{itemize}
    \item The runtime user utterances detected live, based on the Silero voice activity segments and corresponding Whisper speech recognition results;
    \item The manually-transcribed user utterances, based on the manual segmentations and corresponding user audio.
    \item The runtime system utterances, based on the synthesis text and corresponding synthesized audio segments;
\end{itemize}

The word-level timestamps can provide useful information for a variety of streaming tasks and challenges.

\subsection{Task Success}

We manually classified each task execution session into one of four different categories: \emph{correctly-completed}, \emph{incorrectly-completed}, \emph{abandoned} and \emph{system-failure}. A task execution session was labeled \emph{correctly-completed} if the participant reached the correct world state at the end of the task \ie, the same world state the expert reached during the expert demonstration session. Note that it is possible that in correctly-completed sessions some of the sub-steps were done incorrectly, but either redone at a later stage or recovered via some alternative method, in such a way that at the end, when the participant indicated they are done with the last step of the procedure, the correct end state was reached. A task execution session was labeled \emph{incorrectly-completed} if the participant reached the end of the task by advancing through the steps, but did not reach the correct world state at the end of the task; for instance, this can happen if some of the sub-steps are not performed correctly, in a manner that leads to an incorrect result at the end of the task. Sessions where the participant decided to stop before reaching the last step were labeled as \emph{abandoned}. Finally, sessions in which the system failed in a manner that led to an early stop of the interaction (\eg, crashes, significant performance degradation observed by the experimenter, \etc) were labeled as \emph{system-failure}.

We computed \emph{task success rate} by eliminating the \emph{system-failure} sessions and then dividing the number of correctly-completed sessions by the sum of correctly-completed, incorrectly-completed and abandoned sessions. The counts of task execution sessions falling into each of the categories described above and the task success rates are reported on the last column in Table \ref{tab:TaskSuccess}.

\begin{table*}[t]
  \centering
  \begin{tabularx}{\textwidth}{p{4cm}cccccS}
  \toprule
         & Total & Correctly & Incorrectly & Abandoned & System & \multicolumn{1}{c}{Task Success} \\
    Task & Sessions (\#) & Completed (\#) & Completed (\#) & (\#) & Failure (\#) & \multicolumn{1}{c}{Rate (\%)} \\
    \midrule
     Nespresso & 17 & 13 & 0 & 4 & 0 & 76.5 \% \\
     Hard-drive& 15 & 14 & 0 & 0 & 1 & 100.0 \% \\
     Skateboard & 10 & 9 & 0 & 1 & 0 & 90.0 \% \\
     Button & 12 & 7 & 0 & 3 & 2 & 70.0 \% \\
     Notebook & 15 & 7 & 3 & 3 & 2 & 53.8 \% \\
     Mocktail & 16 & 10 & 5 & 1 & 0 & 62.5 \% \\
     \hspace{0.5cm}Margarita & 5 & 5 & 0 & 0 & 0 & 100.0 \% \\
     \hspace{0.5cm}Mojito & 6 & 3 & 2 & 1 & 0 & 50.0 \% \\
     \hspace{0.5cm}Whiskey & 5 & 2 & 3 & 0 & 0 & 40.0 \% \\
    \midrule
    Overall (Dataset) & 85 & 60 & 8 & 12 & 5 & 75.0 \% \\
    \bottomrule
  \end{tabularx}
  \caption{Classification of task execution sessions and task success rates.}
  \label{tab:TaskSuccess}
\end{table*}

\subsection{Gaze Signal Post-Processing}

Gaze behavior is a rich and powerful signal for inferring various aspects of the user's cognitive state, such as confidence, confusion, distraction, frustration, flow state, and so on \cite{stiber2024uh}. To better support such modeling and inference, we performed some additional post-processing on the gaze signal.

First, we generated an annotation stream marking the temporal periods when the participant was likely reading the instructions displayed in the virtual task panel, rather than looking at the physical task space (the intervals are illustrated in Figure \ref{fig:PsiStudioCapture}.K). We generated these annotations automatically by intersecting the user's gaze ray with the location of the floating user interface rectangle holograms in 3D space (see Figure \ref{fig:PsiStudioCapture}.B). To smooth these annotations, gaze-to-interface periods shorter than 0.5 seconds were eliminated, and annotations with intervening gaps of less than 0.5 seconds were merged together.

In addition, we also computed the intersection of the gaze ray with the 3D depth map, then projected that 3D intersection point onto the color image, both (left and right) grayscale image views, and depth image. This gaze point projection was accomplished by leveraging the camera extrinsics and intrinsics information captured in the data. The resulting data streams contain the pixel coordinates for the gaze point in each of the corresponding image spaces.

\section{Conclusion}
\label{sec:Conclusion}

We have introduced \textsc{SigmaCollab}, an interactive, application-driven dataset that aims to support research on various challenges related to real-time collaboration in physically situated settings. The dataset contains approximately 14 hours of interaction data in which untrained participants were guided by a mixed-reality AI assistive application through performing a variety of procedural tasks. \textsc{SigmaCollab} subsumes a rich set of multimodal data streams, including egocentric views from color, grayscale and depth cameras, head, hand and gaze tracking information, as well as additional annotations, such as manual transcripts and word-level timings for the spoken utterances.

We believe that an application-driven approach to dataset construction confers important properties for identifying and studying a variety of challenges related to interaction, coordination and collaboration in physically situated settings. In future work, we plan to use the data to establish a set of new benchmarks that complement preexisting egocentric computer vision benchmarks by more closely focusing on interaction-related challenges, such as timing, proactive interventions, grounding, detecting user cognitive states such as frustration and confusion, \etc We hope the community joins us in adopting this new resource, together with the \textsc{Sigma} open-source application that was used to collect it, and leverages it to make progress towards seamless collaboration between humans and machines in the physical world.

\footnotesize

\normalsize
\bibliographystyle{abbrv-doi}
\bibliography{main}


\end{multicols}  

\newpage
\section*{Appendix}

\setcounter{section}{0}
\setcounter{table}{0}
\setcounter{figure}{0}

\renewcommand{\thesection}{A\arabic{section}}
\renewcommand{\thetable}{A\arabic{table}}
\renewcommand{\thefigure}{A\arabic{figure}}

\section{A Sample Interaction with \textsc{Sigma}}
\label{app:InteractionSample}
Below we present the transcript of an interaction between \textsc{Sigma} and one of the participants, denoted by \textbf{\textcolor{blue!80!black}{User}}.

\begin{tabularx}{\linewidth}{rrX}
\textsubscript{+00:03.9} & \textsc{{Sigma}}: & Hi! This is Sigma, your mixed-reality task assistant. \\
\textsubscript{+00:08.9} & \textsc{{Sigma}}: & Today I'm here to help you make a button using a button press machine. \\
\textsubscript{+00:12.6} & \textsc{{Sigma}}: & You can move me anytime to a more convenient location by pinching and dragging the blue diamond gem. \\
\textsubscript{+00:18.4} & \textsc{{Sigma}}: & Let me know when you're ready to start. \\
\textsubscript{+00:23.7} & \textbf{\textcolor{blue!80!black}{User}}: & \texttt{\textbf{\textcolor{blue!80!black}{i'm ready}}} \\
\textsubscript{+00:25.7} & \textsc{{Sigma}}: & The first step is to setup the cutting system. \\
\textsubscript{+00:28.8} & \textsc{{Sigma}}: & First, attach the white top cutting insert into the top housing. The top cutting insert is the smaller white piece with the notches. Hold it with the notches up and align one of the notches to the arrow on the top housing and press up to magnetize. \\
\textsubscript{+00:45.3} & \textbf{\textcolor{blue!80!black}{User}}: & \texttt{\textbf{\textcolor{blue!80!black}{next step}}} \\
\textsubscript{+00:46.9} & \textsc{{Sigma}}: & Next, place the base cutting insert into the base housing, with the flat face up. The base cutting insert is the bigger white piece. \\
\textsubscript{+00:57.8} & \textbf{\textcolor{blue!80!black}{User}}: & \texttt{\textbf{\textcolor{blue!80!black}{hmm}}} \\
\textsubscript{+01:02.9} & \textbf{\textcolor{blue!80!black}{User}}: & \texttt{\textbf{\textcolor{blue!80!black}{does this look right}}} \\
\textsubscript{+01:07.3} & \textsc{{Sigma}}: & Yes, the base cutting insert looks correctly placed with the flat face up in the base housing. \\
\textsubscript{+01:13.1} & \textbf{\textcolor{blue!80!black}{User}}: & \texttt{\textbf{\textcolor{blue!80!black}{great next step}}} \\
\textsubscript{+01:14.9} & \textsc{{Sigma}}: & Great. \\
\textsubscript{+01:16.0} & \textsc{{Sigma}}: & The next step is to prepare the design for cutting. \\
\textsubscript{+01:19.3} & \textsc{{Sigma}}: & First, place the paper with the design on the white base. \\
\textsubscript{+01:23.9} & \textbf{\textcolor{blue!80!black}{User}}: & \texttt{\textbf{\textcolor{blue!80!black}{hmm oh boy this is very cute}}} \\
\textsubscript{+01:28.0} & \textbf{\textcolor{blue!80!black}{User}}: & \texttt{\textbf{\textcolor{blue!80!black}{um}}} \\
\textsubscript{+01:32.2} & \textbf{\textcolor{blue!80!black}{User}}: & \texttt{\textbf{\textcolor{blue!80!black}{i might choose this one}}} \\
\textsubscript{+01:36.5} & \textbf{\textcolor{blue!80!black}{User}}: & \texttt{\textbf{\textcolor{blue!80!black}{next step}}} \\
\textsubscript{+01:38.3} & \textsc{{Sigma}}: & Next, place the cutting die on top of the paper. The cutting die is the teal piece with a hole in it. Place it over the paper, with the black side facing down, in contact with the paper. \\
\textsubscript{+01:52.8} & \textbf{\textcolor{blue!80!black}{User}}: & \texttt{\textbf{\textcolor{blue!80!black}{should i be trying to center it in any specific way is the inside of the hole what's gonna be cut out}}} \\
\textsubscript{+02:01.9} & \textsc{{Sigma}}: & Yes, you should center the design so that the part visible through the hole in the cutting die is what you want on the button. make sure the cross lines on the cutting die align with the horizontal and vertical axes of your design. \\
\textsubscript{+02:08.3} & \textbf{\textcolor{blue!80!black}{User}}: & \texttt{\textbf{\textcolor{blue!80!black}{do i need to compen-}}} \\
\textsubscript{+02:14.9} & \textbf{\textcolor{blue!80!black}{User}}: & \texttt{\textbf{\textcolor{blue!80!black}{do i need to compensate for the curvature of the button when i pick where i want to center this}}} \\
\textsubscript{+02:23.9} & \textsc{{Sigma}}: & No, you don't need to compensate for the curvature of the button; just center the design as you want it to appear on the flat surface. \\
\textsubscript{+02:31.7} & \textbf{\textcolor{blue!80!black}{User}}: & \texttt{\textbf{\textcolor{blue!80!black}{okay next step}}} \\
\textsubscript{+02:34.1} & \textsc{{Sigma}}: & Next, if you look from the top at the paper, you will see how the button will look like through the center hole of the cutting die. Make sure things are aligned so the design is centered and the cross lines on top of the cutting die correspond to the horizontal and vertical axes of the design. \\
\textsubscript{+02:49.5} & \textbf{\textcolor{blue!80!black}{User}}: & \texttt{\textbf{\textcolor{blue!80!black}{next step}}} \\
\textsubscript{+02:51.2} & \textsc{{Sigma}}: & Next, swivel the base housing with the design and cutting die under the button press by rotating the whole base clockwise around the metal leg. It should click into place. \\
\textsubscript{+03:06.3} & \textbf{\textcolor{blue!80!black}{User}}: & \texttt{\textbf{\textcolor{blue!80!black}{okay}}} \\
\textsubscript{+03:08.5} & \textbf{\textcolor{blue!80!black}{User}}: & \texttt{\textbf{\textcolor{blue!80!black}{next step}}} \\
\textsubscript{+03:10.2} & \textsc{{Sigma}}: & Finally, if the cutting die has slightly shifted position adjust it so it's relatively centered under the press. \\
\textsubscript{+03:20.0} & \textbf{\textcolor{blue!80!black}{User}}: & \texttt{\textbf{\textcolor{blue!80!black}{okay next step}}} \\
\textsubscript{+03:22.6} & \textsc{{Sigma}}: & Great. \\
\textsubscript{+03:23.9} & \textsc{{Sigma}}: & The next step is to cut the design. \\
\textsubscript{+03:26.4} & \textsc{{Sigma}}: & First, push the top handle down and apply pressure to cut the design. You will need to apply significant pressure down, from a standing position, with both hands on the handle, from the top. You should hear or feel a faint click as the design is being cut, but it may not feel obvious. \\
\textsubscript{+03:59.8} & \textbf{\textcolor{blue!80!black}{User}}: & \texttt{\textbf{\textcolor{blue!80!black}{okay next step}}} \\
\end{tabularx}
\newpage

\begin{tabularx}{\linewidth}{rrX}
\textsubscript{+04:01.7} & \textsc{{Sigma}}: & Next, swivel the base housing back out counterclockwise and remove the teal cutting die and design from the base. \\
\textsubscript{+04:15.1} & \textbf{\textcolor{blue!80!black}{User}}: & \texttt{\textbf{\textcolor{blue!80!black}{next step}}} \\
\textsubscript{+04:16.7} & \textsc{{Sigma}}: & Next, take a look at the design, and, if it was not completely cut, use the scissors to finish the circular cut. \\
\textsubscript{+04:24.4} & \textbf{\textcolor{blue!80!black}{User}}: & \texttt{\textbf{\textcolor{blue!80!black}{it looks like i put my design a little too high and i don't have a complete circle i have a flat top should i pick a new one or can i still use this}}} \\
\textsubscript{+04:37.4} & \textsc{{Sigma}}: & If the flat top cuts off an important part of the design, it's best to pick a new one and try again, ensuring the design is centered properly. \\
\textsubscript{+04:47.1} & \textbf{\textcolor{blue!80!black}{User}}: & \texttt{\textbf{\textcolor{blue!80!black}{it isn't an important part}}} \\
\textsubscript{+04:51.6} & \textsc{{Sigma}}: & Ready for the next step? \\
\textsubscript{+04:53.6} & \textbf{\textcolor{blue!80!black}{User}}: & \texttt{\textbf{\textcolor{blue!80!black}{sure let's do it}}} \\
\textsubscript{+04:55.6} & \textsc{{Sigma}}: & Next, remove the white base cutting insert from the base. \\
\textsubscript{+05:00.5} & \textbf{\textcolor{blue!80!black}{User}}: & \texttt{\textbf{\textcolor{blue!80!black}{next step}}} \\
\textsubscript{+05:02.1} & \textsc{{Sigma}}: & Finally, remove the white top cutting insert from the press head. \\
\textsubscript{+05:07.7} & \textbf{\textcolor{blue!80!black}{User}}: & \texttt{\textbf{\textcolor{blue!80!black}{next step}}} \\
\textsubscript{+05:09.5} & \textsc{{Sigma}}: & Great. \\
\textsubscript{+05:10.7} & \textsc{{Sigma}}: & The next step is to setup the button press. \\
\textsubscript{+05:13.8} & \textsc{{Sigma}}: & First, attach the top insert to the top housing. The top insert is the teal cylindrical piece with small notches. Hold it with the notches up, align the notch marked with the letter `B` with the arrow on the top housing and press upward to magnetize. The outer ring will  \\
\textsubscript{+05:29.7} & \textbf{\textcolor{blue!80!black}{User}}: & \texttt{\textbf{\textcolor{blue!80!black}{hmm}}} \\
\textsubscript{+05:44.4} & \textbf{\textcolor{blue!80!black}{User}}: & \texttt{\textbf{\textcolor{blue!80!black}{hmm}}} \\
\textsubscript{+05:50.7} & \textbf{\textcolor{blue!80!black}{User}}: & \texttt{\textbf{\textcolor{blue!80!black}{okay next step}}} \\
\textsubscript{+05:53.1} & \textsc{{Sigma}}: & Next, place the base insert on the base housing and rotate until the piece clicks and locks into place. The base insert is the teal cylindrical piece with large notches. \\
\textsubscript{+06:11.1} & \textbf{\textcolor{blue!80!black}{User}}: & \texttt{\textbf{\textcolor{blue!80!black}{next step}}} \\
\textsubscript{+06:12.8} & \textsc{{Sigma}}: & Great. \\
\textsubscript{+06:14.0} & \textsc{{Sigma}}: & The next step is to load the button into the press. \\
\textsubscript{+06:17.2} & \textsc{{Sigma}}: & First, rotate both top and base inserts into position marked with the letter `A`, aligning with arrows on the top and bottom housing. \\
\textsubscript{+06:25.9} & \textbf{\textcolor{blue!80!black}{User}}: & \texttt{\textbf{\textcolor{blue!80!black}{next step}}} \\
\textsubscript{+06:27.6} & \textsc{{Sigma}}: & Next, place a rounded metal button shell with the rounded face up into the base insert. \\
\textsubscript{+06:45.9} & \textbf{\textcolor{blue!80!black}{User}}: & \texttt{\textbf{\textcolor{blue!80!black}{next step}}} \\
\textsubscript{+06:47.6} & \textsc{{Sigma}}: & Next, layer the cut paper design over the button shell. Use the pin line as a guide to make sure the text and imagery is straight. \\
\textsubscript{+07:03.5} & \textbf{\textcolor{blue!80!black}{User}}: & \texttt{\textbf{\textcolor{blue!80!black}{next step}}} \\
\textsubscript{+07:05.3} & \textsc{{Sigma}}: & Next, take a mylar piece out of the teal envelope and place it over the cut paper design. \\
\textsubscript{+07:23.2} & \textbf{\textcolor{blue!80!black}{User}}: & \texttt{\textbf{\textcolor{blue!80!black}{next step}}} \\
\textsubscript{+07:24.8} & \textsc{{Sigma}}: & Next, swivel the base housing into place below the button press by rotating clockwise. It should click into place. \\
\textsubscript{+07:40.8} & \textbf{\textcolor{blue!80!black}{User}}: & \texttt{\textbf{\textcolor{blue!80!black}{next step}}} \\
\textsubscript{+07:42.5} & \textsc{{Sigma}}: & Next, press the handle once. This action will capture the button in the top insert. \\
\textsubscript{+07:49.7} & \textbf{\textcolor{blue!80!black}{User}}: & \texttt{\textbf{\textcolor{blue!80!black}{do i need to press hard}}} \\
\textsubscript{+07:54.3} & \textbf{\textcolor{blue!80!black}{User}}: & \texttt{\textbf{\textcolor{blue!80!black}{oh next step}}} \\
\textsubscript{+07:57.7} & \textsc{{Sigma}}: & Finally, swivel the base back out, counterclockwise. Note: the base should be empty now, the top pieces of the button are inside the top insert. \\
\textsubscript{+08:07.7} & \textbf{\textcolor{blue!80!black}{User}}: & \texttt{\textbf{\textcolor{blue!80!black}{next step}}} \\
\textsubscript{+08:09.6} & \textsc{{Sigma}}: & Great. \\
\textsubscript{+08:10.8} & \textsc{{Sigma}}: & The next step is to make the button. \\
\textsubscript{+08:13.1} & \textsc{{Sigma}}: & First, rotate both the top and base inserts to the position marked with the letter `B`, aligning with the corresponding arrows on the top and bottom housing. \\
\textsubscript{+08:22.6} & \textbf{\textcolor{blue!80!black}{User}}: & \texttt{\textbf{\textcolor{blue!80!black}{next step}}} \\
\textsubscript{+08:26.0} & \textsc{{Sigma}}: & Next, place the metal pin backing with the pin side down and the zig-zag part up into the base insert. \\
\textsubscript{+08:38.7} & \textbf{\textcolor{blue!80!black}{User}}: & \texttt{\textbf{\textcolor{blue!80!black}{next step}}} \\
\textsubscript{+08:40.8} & \textsc{{Sigma}}: & Next, align the pin parallel to the pin line and check that the zig-zag part of the pin is sitting down flush with the metal. \\
\textsubscript{+09:02.3} & \textbf{\textcolor{blue!80!black}{User}}: & \texttt{\textbf{\textcolor{blue!80!black}{next step}}} \\
\textsubscript{+09:03.8} & \textsc{{Sigma}}: & Next, swivel the base housing into place below the button press by rotating clockwise. It should click into place. \\
\end{tabularx}
\newpage

\begin{tabularx}{\linewidth}{rrX}

\textsubscript{+09:18.0} & \textbf{\textcolor{blue!80!black}{User}}: & \texttt{\textbf{\textcolor{blue!80!black}{next step}}} \\
\textsubscript{+09:20.1} & \textsc{{Sigma}}: & Next, press the handle. \\
\textsubscript{+09:29.4} & \textbf{\textcolor{blue!80!black}{User}}: & \texttt{\textbf{\textcolor{blue!80!black}{next step}}} \\
\textsubscript{+09:31.2} & \textsc{{Sigma}}: & Finally, swivel the base housing back out by rotating counterclockwise, and retrieve the finished button. \\
\textsubscript{+09:41.7} & \textbf{\textcolor{blue!80!black}{User}}: & \texttt{\textbf{\textcolor{blue!80!black}{next step}}} \\
\textsubscript{+09:43.6} & \textsc{{Sigma}}: & Congratulations. You have finished the task. \\
\textsubscript{+09:47.7} & \textsc{{Sigma}}: & I'll see you next time. Bye bye. \\
\textsubscript{+09:49.6} & \textbf{\textcolor{blue!80!black}{User}}: & \texttt{\textbf{\textcolor{blue!80!black}{thank you}}} \\
\end{tabularx}

\newpage

\section{Response-Generating Prompt}
\label{app:PromptSample}
Below is a sample response-generating prompt that \textsc{Sigma} used at the end of the \textit{``Does this look great?''} runtime-detected user utterance\footnote{Note that the actual utterance spoken by the user, as shown in Appendix \ref{app:InteractionSample} is ``Does this look right?''; in this case the system committed a speech-recognition error at runtime.} in the sample interaction illustrated in Figure \ref{fig:InteractionSample} and Appendix \ref{app:InteractionSample}.

The prompt template is shown in regular black typeface, and various pieces of context that are automatically populated in the prompt template are shown in different colors: 
\begin{itemize}
    \item \textcolor{blue}{blue}: the task name
    \item \textcolor{green!50!black}{dark green}: the task recipe
    \item \textcolor{cyan!50!black}{dark cyan}: the current step the user is on
    \item \textcolor{brown}{brown}: the current sub-step the user is on
    \item \textcolor{red!50!yellow}{orange}: the history of the dialog in the sub-step
    \item \textcolor{red!50!blue}{purple}: the last speech recognition result
\end{itemize}

In addition, an image collected from the right grayscale camera at the time the user started speaking their utterance is also injected into the prompt template, as shown below. 

\textsc{Prompt:}

\begin{adjustwidth}{1cm}{1cm}

{\obeylines
You are a helpful assistant guiding a user via spoken interaction to perform the following task in the physical world: \textcolor{blue}{Make a button using a button press machine}

You are helping the user to perform this task based on the following task recipe: 
\textcolor{green!50!black}{Make a button using a button press machine
1. Setup the cutting system.
  1.1. Attach the white <b>top cutting insert</b> into the <b>top housing</b>. The <b>top cutting insert</b> is the smaller white piece with the notches. Hold it with the notches up and align one of the notches to the arrow on the <b>top housing</b> and press up to magnetize.
  1.2. Place the <b>base cutting insert</b> into the <b>base housing</b>, with the flat face up. The <b>base cutting insert</b> is the bigger white piece.

2. Prepare the design for cutting.
  2.1. Place the paper with the design on the white base.
  2.2. Place the <b>cutting die</b> on top of the paper. The <b>cutting die</b> is the teal piece with a hole in it. Place it over the paper, with the black side facing down, in contact with the paper.

[...]\footnote{the reminder of the recipe is omitted for space reasons; the full task recipe is available in Appendix \ref{app:TaskRecipes}}
}

The user is on the following step: \textcolor{cyan!50!black}{1. Setup the cutting system}
The user is on the following substep: \textcolor{brown}{1.2. Place the <b>base cutting insert</b> into the <b>base housing</b>, with the flat face up. The <b>base cutting insert</b> is the bigger white piece.}

Here is a transcript of the dialog since the instructions for the last substep were delivered to the user, where the timings of each utterance are shown at the beginning of each utterance, in seconds, relative to time 0, when the instruction was first delivered:
\textcolor{red!50!yellow}{[+00:00.0] You(Assistant): Next, place the base cutting insert into the base housing, with the flat face up. The base cutting insert is the bigger white piece.
[+00:11.1] User:  Hmm.
[+00:16.2] User:  Does this look great?}

The speech recognition result for the last user utterance was: \textcolor{red!50!blue}{" Does this look great?"}

The utterance spoken by the user is probably close to this speech recognition results, but there might be some slight differences because of speech recognition errors.

For additional context, here is also the image that the user saw when they started speaking their utterance: 

\includegraphics[width=0.3\linewidth]{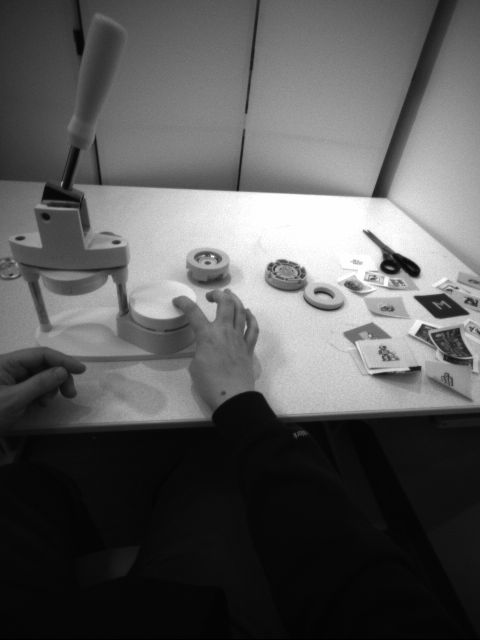}

Based on the dialog context and speech recognition results, and using information from image as context, please classify the LAST user utterance in one of four classes: question, no-question, next-step, or step-navigation. If you think the utterance is a question, also provide an appropriate answer. 

Please provide your response using precisely the following json format:
\{
\hspace{1cm}{"classification": "question"},
\hspace{1cm}{"answer": "The apple is on the red box"}
\}

Here is more information on how to perform the classification: 

- If the user utterance is a question related to this step or substep, then classify the utterance as "question", place your answer to the question in the answer field. Please provide a concise answer, as the answer will be spoken back to the user.

- If the user utterance indicates that they want to move to the next substep, or if it describes a state of affairs where they have completed the substep then classify as "next-step" and leave the answer field blank. 

- If the user is issuing a navigational command to go to a different step, substep or step.substep combination, or if they are signaling they are already on a different step or substep, then classify the utterance as "step-navigation" and provide the target step and substep in the form "step=\#,substep=\#" in the answer. If only a step number is provided by the user, answer in the form "step=\#". 

- If the user looks like they are talking or making a comment to themselves, or if it looks like they are repeating the instruction (or a portion of it) to themselves, then classify as "no-question" and leave the answer field blank. 

- If you are unsure or in all other cases, please classify the utterance as "no-question" and leave the answer field blank. 

Provide your response exactly matching the JSON format described above and make sure you provide a well-formatted json. Do not provide any text before or after the json, just start with an open brace and end with a closing brace.
}
\end{adjustwidth}

\textsc{Model Response:}

\begin{adjustwidth}{1cm}{1cm}

{\obeylines
\{
\hspace{1cm}"classification": "question",
\hspace{1cm}"answer": "Yes, the base cutting insert looks correctly placed with the flat face up in the base housing."
\}
}

\end{adjustwidth}

\newpage

\section{Dataset Sections}
\label{app:DatasetSections}

Three different Azure GPT model deployments were used in conjunction with the response-generating prompt throughout the data collection study. In the first part of the study we relied on a GPT-4o model using a global data zone region. After observing relatively large response latencies, we switched to a GPT-4o-mini deployment in a US data zone region. Finally, for the last third of the dataset, we switched back to a GPT-4o deployment using a US data zone region.

Because of these changes, these three sections of the dataset --- which we refer to as Section A, B, and C --- have slightly different characteristics in terms of latency and GPT responses to user queries. Table \ref{tab:DatasetSections} provides the details regarding which task execution sessions belong in each section, and Table \ref{tab:TaskTypesBySection} shows the number of task execution sessions for each task type across the three dataset sections.

\begin{table*}[h]
  \centering
  \begin{tabular}{clccc}
  \toprule
    Section & GPT model & From Session & To Session & Sessions Count \\
    \midrule
    A & GPT-4o/Global Data Zone & 20250228-134800 & 20250307-102820 & 28 \\
    B & GPT-4o-mini/US Data Zone & 20250318-154110 & 20250324-105658 & 27 \\
    C & GPT-4o/US Data Zone & 20250325-110228 & 20250417-160112 & 30 \\
    \bottomrule
  \end{tabular}
  \caption{Dataset Sections.}
  \label{tab:DatasetSections}
\end{table*}

\begin{table*}[h]
  \centering
  \begin{tabular}{p{4cm}cccc}
  \toprule
         & Sessions & Sessions & Sessions & Sessions \\
    Task & in Section A (\#) & in Section B (\#) & in Section C (\#) & Total (\#) \\
    \midrule
     Nespresso & 6 & 5 & 6 & 17 \\
     Hard-drive& 5 & 6 & 4 & 15 \\
     Skateboard & 3 & 3 & 4 & 10 \\
     Button & 5 & 3 & 4 & 12 \\
     Notebook & 5 & 5 & 5 & 15 \\
     Mocktail & 4 & 5 & 7 & 16 \\
     \hspace{0.5cm}Margarita & 2 & 2 & 1 & 5 \\
     \hspace{0.5cm}{Mojito} & 1 & 2 & 3 & 6 \\
     \hspace{0.5cm}{Whiskey} & 1 & 1 & 3 & 5 \\
    \midrule
    Overall (Dataset) & 85 & 60 & 8 & 12 \\
    \bottomrule
  \end{tabular}
  \caption{Task execution counts by task type in each dataset section.}
  \label{tab:TaskTypesBySection}
\end{table*}

\newpage

\section{Task Recipes}
\label{app:TaskRecipes}
Below are the recipes for the 8 tasks used in the \textsc{SigmaCollab} dataset. For each task, four images sampled from a reference execution of the task are also shown for illustration purposes. The classification of the sub-step per the taxonomy described in Section \ref{subsec:SigmaTasks} is also shown after each sub-step instructions, in bracketed light gray text.

\textbf{Task 1: Make a Mojito mocktail}
\begin{figure}[h!]
    \centering
    \setlength{\fboxsep}{0pt}
    \fbox{\includegraphics[width=4.4cm]{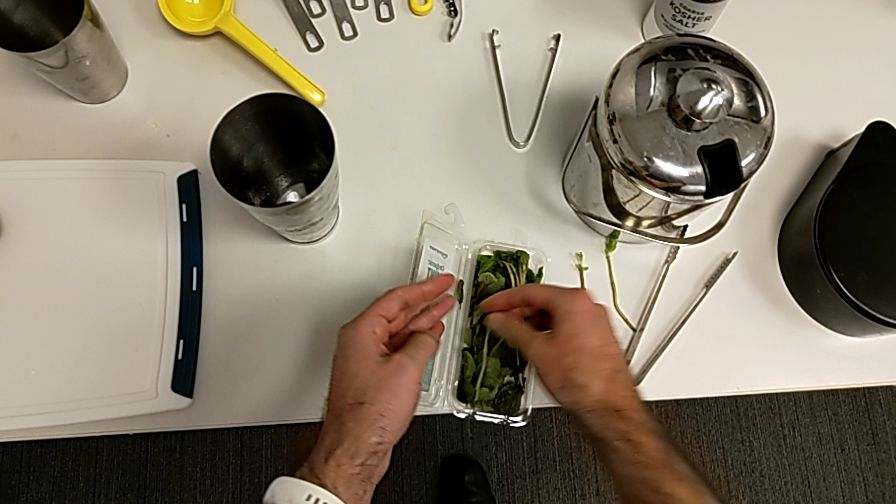}}
    \fbox{\includegraphics[width=4.4cm]{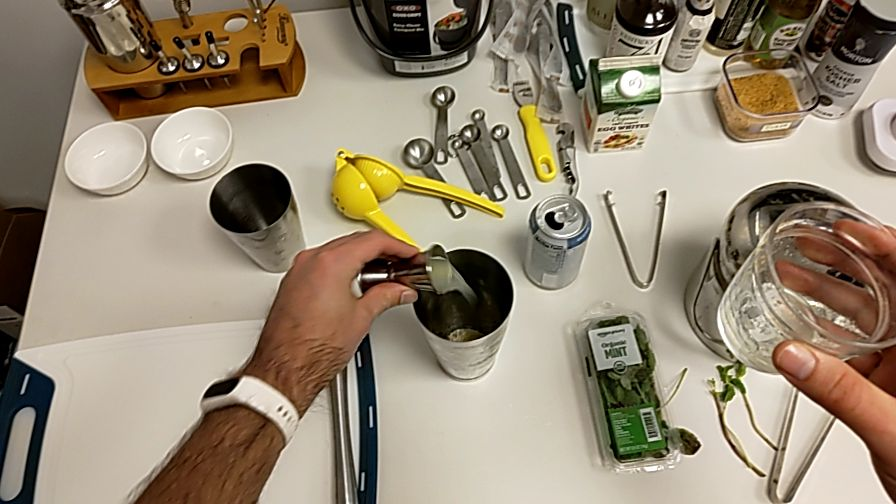}}
    \fbox{\includegraphics[width=4.4cm]{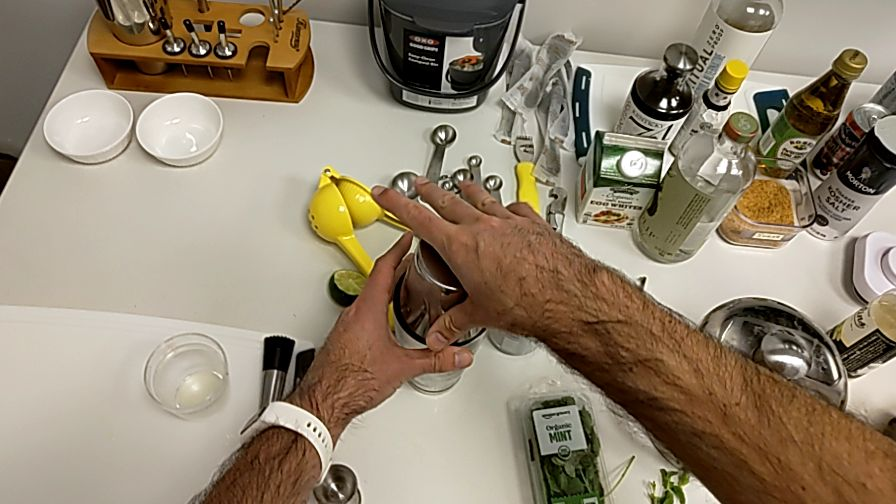}}
    \fbox{\includegraphics[width=4.4cm]{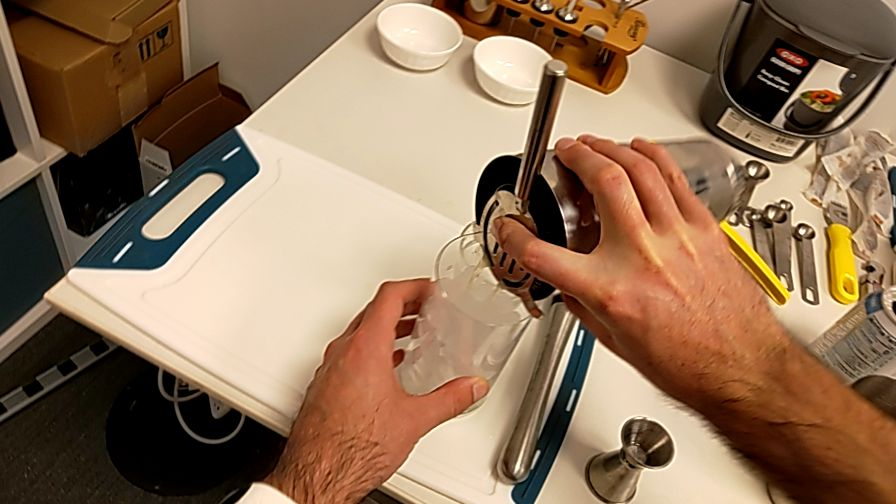}}
\end{figure}
\begin{enumerate}[label*=\arabic*.]
	\item Muddle the mint
	\begin{enumerate}[label*=\arabic*.]
		\item Add 7 mint leaves in the larger tin of a Boston shaker. \textcolor{lightgray}{[Add Amount]}
		\item Add a teaspoon of soda water. \textcolor{lightgray}{[Add Amount]}
		\item Add half a teaspoon of sugar. \textcolor{lightgray}{[Add Amount]}
		\item Muddle the mint by repeatedly hammering it in the bottom of the tin with the flat end of the muddler. \textcolor{lightgray}{[Invisible State]}
	\end{enumerate}
	\item Prepare the ingredients
	\begin{enumerate}[label*=\arabic*.]
		\item Using the citrus squeezer, extract the juice from two lime halfs into a cup. \textcolor{lightgray}{[Add Amount]}
		\item Using the jigger, measure and pour 3/4 ounces of fresh lime juice into the tin with the mint. \textcolor{lightgray}{[Add Amount]}
		\item Pour 2 ounces of Seedlip Garden non-alcoholic gin into the tin. \textcolor{lightgray}{[Add Amount]}
		\item Pour in half an ounce of simple syrup. \textcolor{lightgray}{[Add Amount]}
		\item Add 6 or 7 ice-cubes into the tin. \textcolor{lightgray}{[Add Amount]}
	\end{enumerate}
	\item Shake the cocktail
	\begin{enumerate}[label*=\arabic*.]
		\item Cover the large tin of the Boston shaker with the smaller tin, and push the top (smaller) tin down firmly to ensure a seal. \textcolor{lightgray}{[Invisible State]}
		\item Using both hands to catch the tins, make sure they stay pressed together and shake the drink 7 to ten seconds, until the tins become cold to the touch. Make sure you hold them tight, as the shaking process might have them come undone and the drink might spill. \textcolor{lightgray}{[Invisible State]}
		\item Open the top tin. If it does not come out easily, you can tap it from the side, or try to rotate it up. \textcolor{lightgray}{[Move]}
		\item Place a Hawthorne strainer over the opening of the large tin. \textcolor{lightgray}{[Move]}
	\end{enumerate}
	\item Pour the mocktail
	\begin{enumerate}[label*=\arabic*.]
		\item Add a mint sprig into a highball glass. \textcolor{lightgray}{[Discrete Event]}
		\item Fill the glass three quarters with ice. \textcolor{lightgray}{[Add Amount]}
		\item While holding the strainer over the top of the tin, tilt the tin and strain the drink into the prepared glass. \textcolor{lightgray}{[Add Amount]}
		\item Top the glass with soda water. \textcolor{lightgray}{[Add Amount]}
	\end{enumerate}
\end{enumerate}
\newpage
\textbf{Task 2: Make coffee using a Nespresso Pixie machine}
\begin{figure}[h!]
    \centering
    \setlength{\fboxsep}{0pt}
    \fbox{\includegraphics[width=4.4cm]{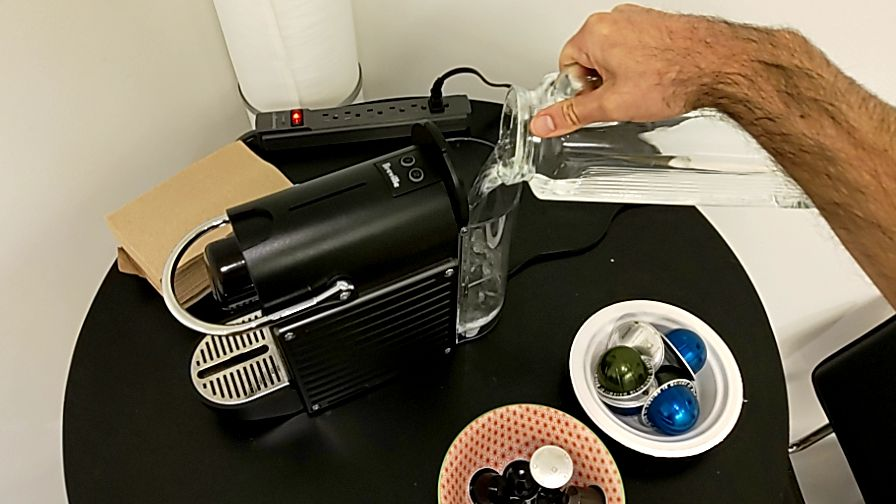}}
    \fbox{\includegraphics[width=4.4cm]{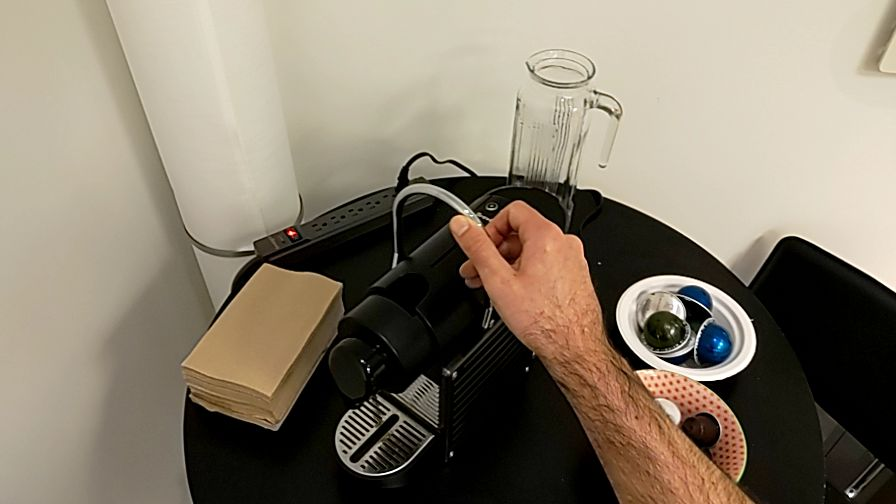}}
    \fbox{\includegraphics[width=4.4cm]{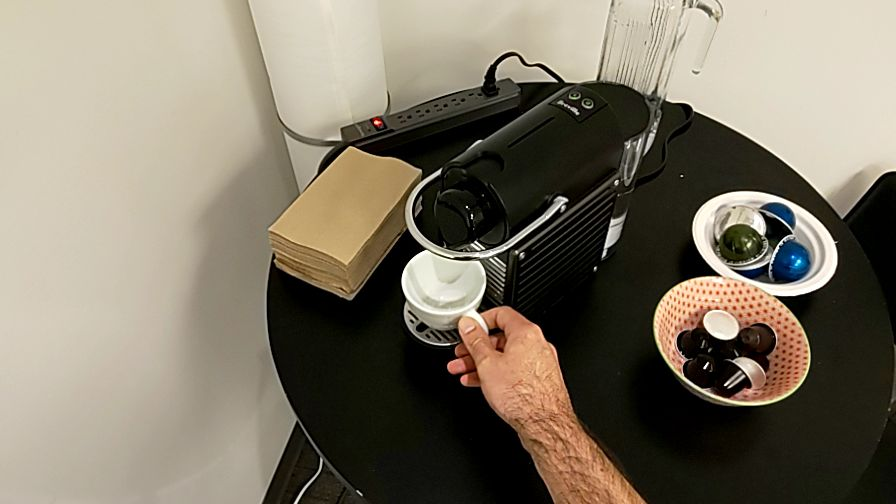}}
    \fbox{\includegraphics[width=4.4cm]{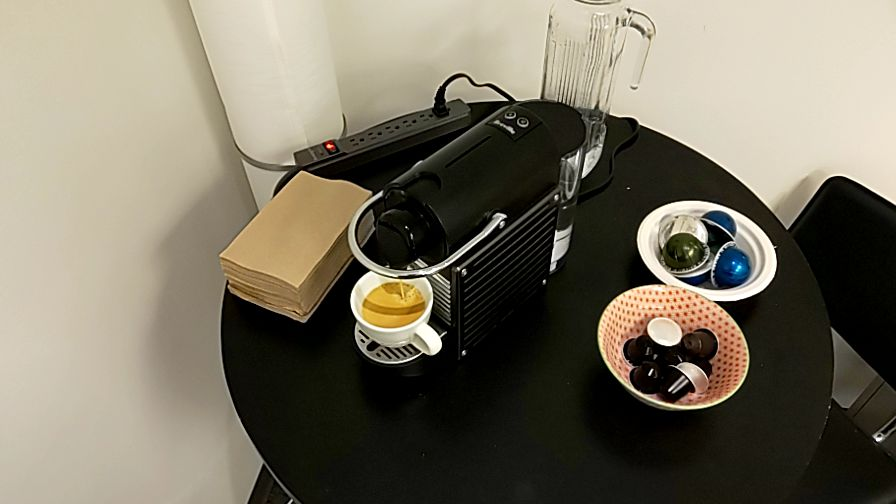}}
\end{figure}
\begin{enumerate}[label*=\arabic*.]
	\item Fill the water tank with fresh water
	\begin{enumerate}[label*=\arabic*.]
		\item Open the water tank lid \textcolor{lightgray}{[Discrete Event]}
		\item Fill the water tank with fresh water \textcolor{lightgray}{[Add Amount]}
		\item Close the water tank lid \textcolor{lightgray}{[Discrete Event]}
	\end{enumerate}
	\item Turn on the machine
	\begin{enumerate}[label*=\arabic*.]
		\item Briefly push the two buttons on top of the machine simultaneously. The buttons should start blinking. \textcolor{lightgray}{[Discrete Event]}
		\item Wait until the buttons stop blinking and the machine is ready for use \textcolor{lightgray}{[Discrete Event]}
	\end{enumerate}
	\item Insert a coffee capsule
	\begin{enumerate}[label*=\arabic*.]
		\item Lift the front lever upwards. \textcolor{lightgray}{[Discrete Event]}
		\item Insert a Nespresso coffee capsule into the capsule compartment, with the flat side towards the front of the machine. \textcolor{lightgray}{[Discrete Event]}
		\item Lower the lever to close the compartment. \textcolor{lightgray}{[Discrete Event]}
	\end{enumerate}
	\item Brew the coffee
	\begin{enumerate}[label*=\arabic*.]
		\item Place the coffee cup on the drip tray below the nozzle. \textcolor{lightgray}{[Move]}
		\item Press the espresso or the lungo button and wait until the coffee stops brewing. \textcolor{lightgray}{[Discrete Event]}
	\end{enumerate}
\end{enumerate}
\newpage
\textbf{Task 3: Replace a hard-drive in a PC}
\begin{figure}[h!]
    \centering
    \setlength{\fboxsep}{0pt}
    \fbox{\includegraphics[width=4.4cm]{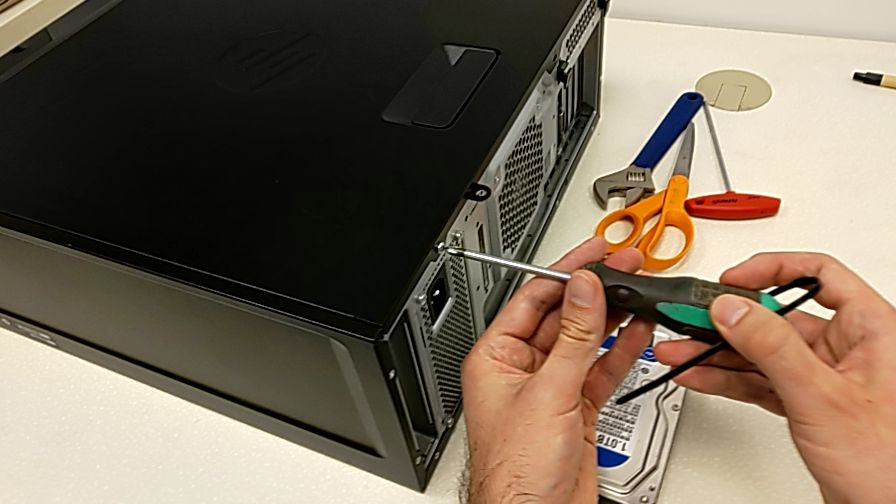}}
    \fbox{\includegraphics[width=4.4cm]{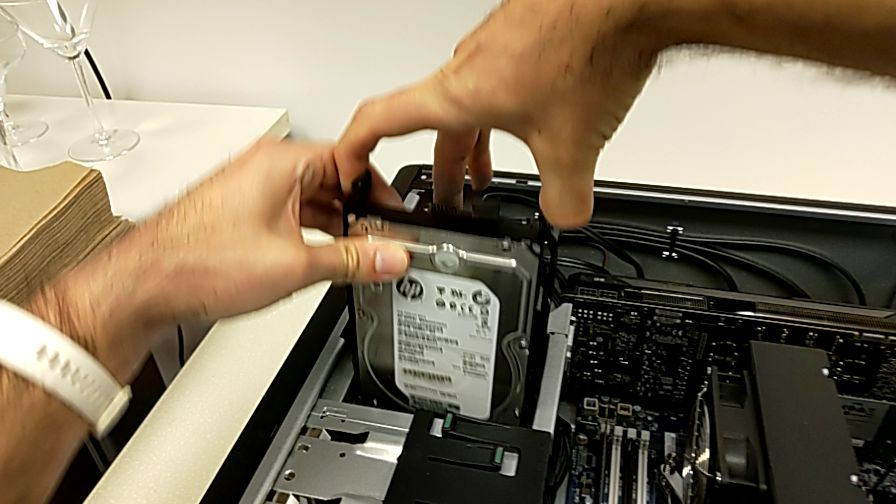}}
    \fbox{\includegraphics[width=4.4cm]{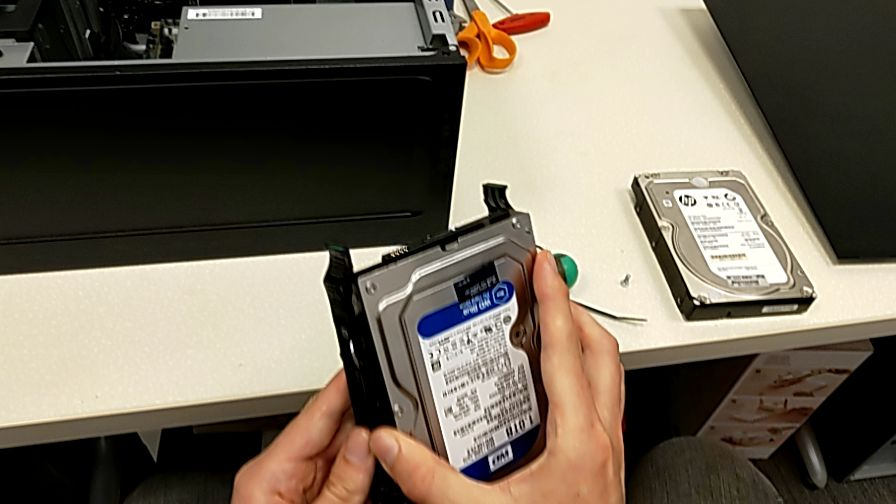}}
    \fbox{\includegraphics[width=4.4cm]{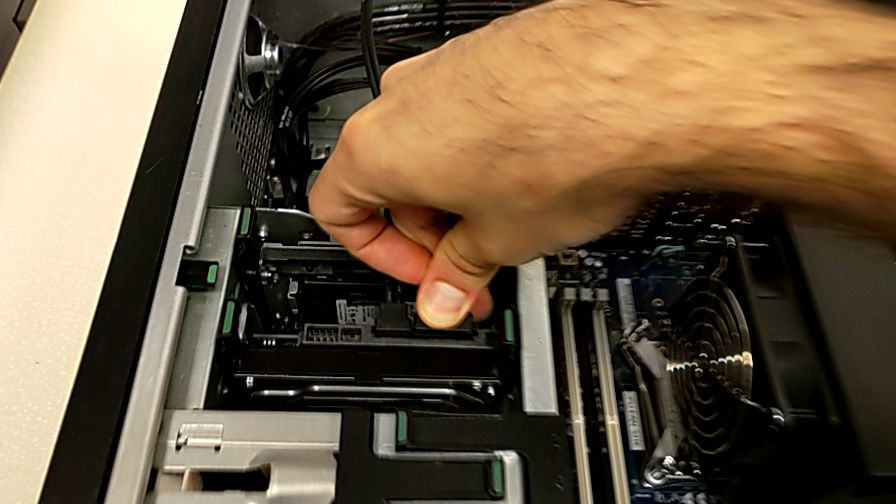}}
\end{figure}
\begin{enumerate}[label*=\arabic*.]
	\item Open the PC case
	\begin{enumerate}[label*=\arabic*.]
		\item To remove the case lid you will have to first unscrew the case screw using the screwdriver. The case screw is the single screw connecting the back of the PC to the black case lid. \textcolor{lightgray}{[Move]}
		\item Pull the case-lid handle and open and remove the case lid. Place the lid on the table next to you. \textcolor{lightgray}{[Move]}
	\end{enumerate}
	\item Remove the existing hard-drive
	\begin{enumerate}[label*=\arabic*.]
		\item Look for two hard drives installed towards the front of the PC case. Disconnect the wide connector marked D2, from the hard drive that is closer to you. \textcolor{lightgray}{[Move]}
		\item Disconnect the narrow connector. \textcolor{lightgray}{[Move]}
		\item Lift the hard-drive out by pushing the two green tabs towards each other and pulling up. \textcolor{lightgray}{[Move]}
		\item Remove the plastic casing from around the hard-drive. \textcolor{lightgray}{[Move]}
	\end{enumerate}
	\item Insert the new hard-drive
	\begin{enumerate}[label*=\arabic*.]
		\item Place the plastic casing around the new hard-drive. \textcolor{lightgray}{[Discrete Event]}
		\item Insert the hard-drive into the bay. \textcolor{lightgray}{[Discrete Event]}
		\item Plug the narrow connector into the hard-drive. \textcolor{lightgray}{[Discrete Event]}
		\item Plug the wide connector into the hard-drive. \textcolor{lightgray}{[Discrete Event]}
	\end{enumerate}
	\item Close the PC case
	\begin{enumerate}[label*=\arabic*.]
		\item Slide the case lid in and close it down. \textcolor{lightgray}{[Discrete Event]}
		\item Screw the case screw back using the screwdriver. \textcolor{lightgray}{[Invisible State]}
	\end{enumerate}
\end{enumerate}
\newpage
\textbf{Task 4: Make a button using a button press machine}
\begin{figure}[h!]
    \centering
    \setlength{\fboxsep}{0pt}
    \fbox{\includegraphics[width=4.4cm]{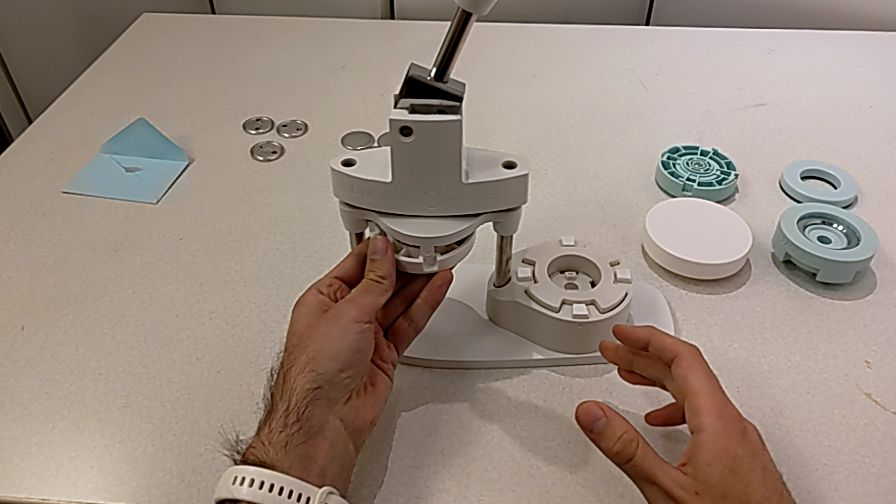}}
    \fbox{\includegraphics[width=4.4cm]{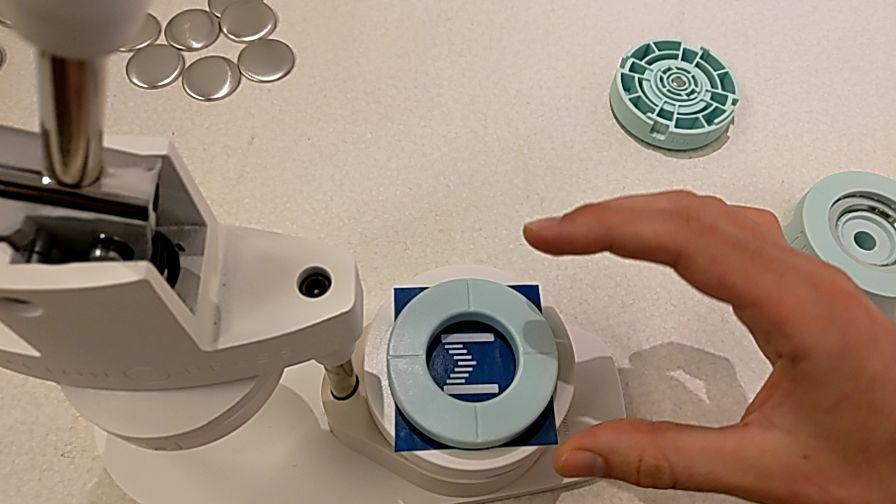}}
    \fbox{\includegraphics[width=4.4cm]{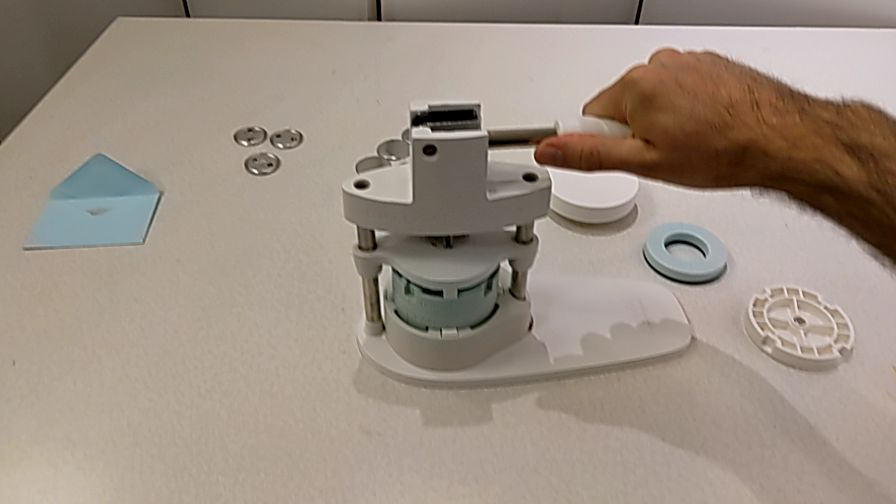}}
    \fbox{\includegraphics[width=4.4cm]{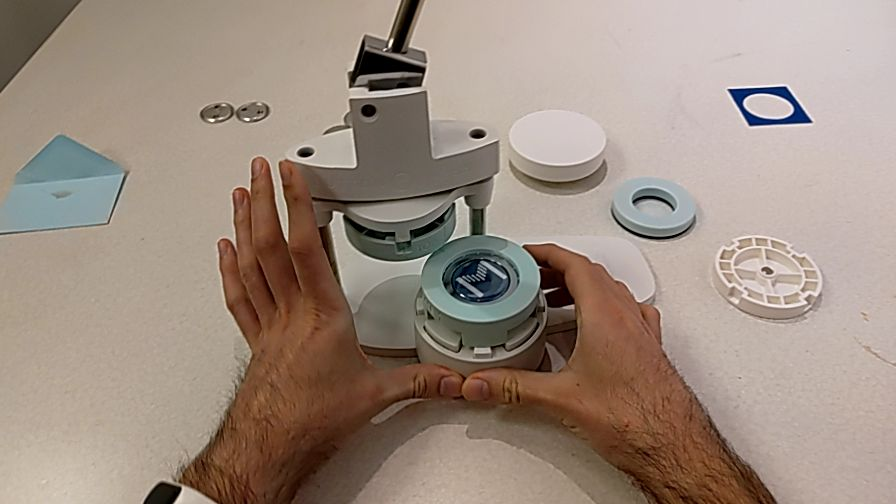}}
\end{figure}
\begin{enumerate}[label*=\arabic*.]
	\item Setup the cutting system
	\begin{enumerate}[label*=\arabic*.]
		\item Attach the white top cutting insert into the top housing. The top cutting insert is the smaller white piece with the notches. Hold it with the notches up and align one of the notches to the arrow on the top housing and press up to magnetize. \textcolor{lightgray}{[Discrete Event]}
		\item Place the base cutting insert into the base housing, with the flat face up. The base cutting insert is the bigger white piece. \textcolor{lightgray}{[Discrete Event]}
	\end{enumerate}
	\item Prepare the design for cutting
	\begin{enumerate}[label*=\arabic*.]
		\item Place the paper with the design on the white base. \textcolor{lightgray}{[Move]}
		\item Place the cutting die on top of the paper. The cutting die is the teal piece with a hole in it. Place it over the paper, with the black side facing down, in contact with the paper. \textcolor{lightgray}{[Move]}
		\item If you look from the top at the paper, you will see how the button will look like through the center hole of the cutting die. Make sure things are aligned so the design is centered and the cross lines on top of the cutting die correspond to the horizontal and vertical axes of the design. \textcolor{lightgray}{[Check]}
		\item Swivel the base housing with the design and cutting die under the button press by rotating the whole base clockwise around the metal leg. It should click into place. \textcolor{lightgray}{[Discrete Event]}
		\item If the cutting die has slightly shifted position adjust it so it's relatively centered under the press. \textcolor{lightgray}{[Check]}
	\end{enumerate}
	\item Cut the design
	\begin{enumerate}[label*=\arabic*.]
		\item Push the top handle down and apply pressure to cut the design. You will need to apply significant pressure down, from a standing position, with both hands on the handle, from the top. You should hear or feel a faint click as the design is being cut, but it may not feel obvious. \textcolor{lightgray}{[Invisible State]}
		\item Swivel the base housing back out counterclockwise and remove the teal cutting die and design from the base. \textcolor{lightgray}{[Move]}
		\item Take a look at the design, and, if it was not completely cut, use the scissors to finish the circular cut. \textcolor{lightgray}{[Check]}
		\item Remove the white base cutting insert from the base. \textcolor{lightgray}{[Move]}
		\item Remove the white top cutting insert from the press head. \textcolor{lightgray}{[Move]}
	\end{enumerate}
	\item Setup the button press
	\begin{enumerate}[label*=\arabic*.]
		\item Attach the top insert to the top housing. The top insert is the teal cylindrical piece with small notches. Hold it with the notches up, align the notch marked with the letter `B` with the arrow on the top housing and press upward to magnetize. The outer ring will drop when the center is secure. \textcolor{lightgray}{[Discrete Event]}
		\item Place the base insert on the base housing and rotate until the piece clicks and locks into place. The base insert is the teal cylindrical piece with large notches. \textcolor{lightgray}{[Discrete Event]}
	\end{enumerate}
	\item Load the button into the press
	\begin{enumerate}[label*=\arabic*.]
		\item Rotate both top and base inserts into position marked with the letter `A`, aligning with arrows on the top and bottom housing. \textcolor{lightgray}{[Discrete Event]}
		\item Place a rounded metal button shell with the rounded face up into the base insert. \textcolor{lightgray}{[Move]}
		\item Layer the cut paper design over the button shell. Use the pin line as a guide to make sure the text and imagery is straight. \textcolor{lightgray}{[Move]}
		\item Take a mylar piece out of the teal envelope and place it over the cut paper design. \textcolor{lightgray}{[Move]}
		\item Swivel the base housing into place below the button press by rotating clockwise. It should click into place. \textcolor{lightgray}{[Discrete Event]}
		\item Press the handle once. This action will capture the button in the top insert. \textcolor{lightgray}{[Discrete Event]}
		\item Swivel the base back out, counterclockwise. Note: the base should be empty now, the top pieces of the button are inside the top insert. \textcolor{lightgray}{[Move]}
	\end{enumerate}
	\item Make the button
	\begin{enumerate}[label*=\arabic*.]
		\item Rotate both the top and base inserts to the position marked with the letter `B`, aligning with the corresponding arrows on the top and bottom housing. \textcolor{lightgray}{[Discrete Event]}
		\item Place the metal pin backing with the pin side down and the zig-zag part up into the base insert. \textcolor{lightgray}{[Move]}
		\item Align the pin parallel to the pin line and check that the zig-zag part of the pin is sitting down flush with the metal. \textcolor{lightgray}{[Check]}
		\item Swivel the base housing into place below the button press by rotating clockwise. It should click into place. \textcolor{lightgray}{[Discrete Event]}
		\item Press the handle. \textcolor{lightgray}{[Discrete Event]}
		\item Swivel the base housing back out by rotating counterclockwise, and retrieve the finished button. \textcolor{lightgray}{[Move]}
	\end{enumerate}
\end{enumerate}
\newpage
\textbf{Task 5: Make a notebook using a binding machine}
\begin{figure}[h!]
    \centering
    \setlength{\fboxsep}{0pt}
    \fbox{\includegraphics[width=4.4cm]{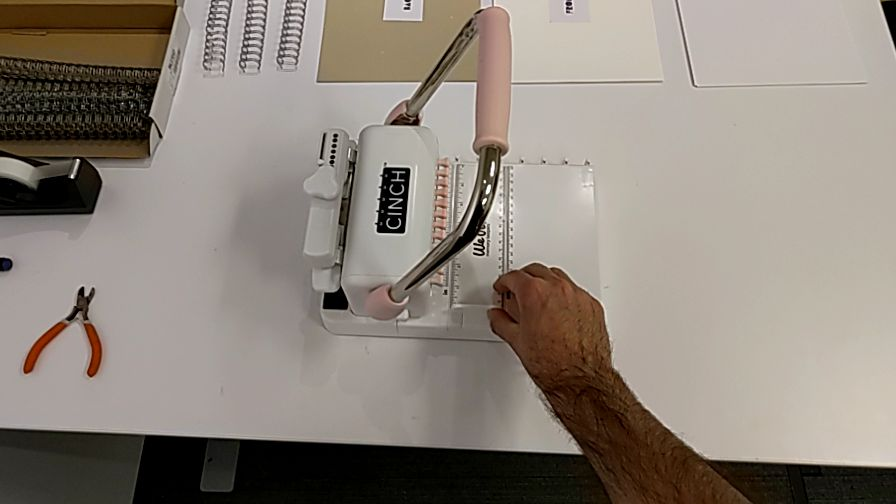}}
    \fbox{\includegraphics[width=4.4cm]{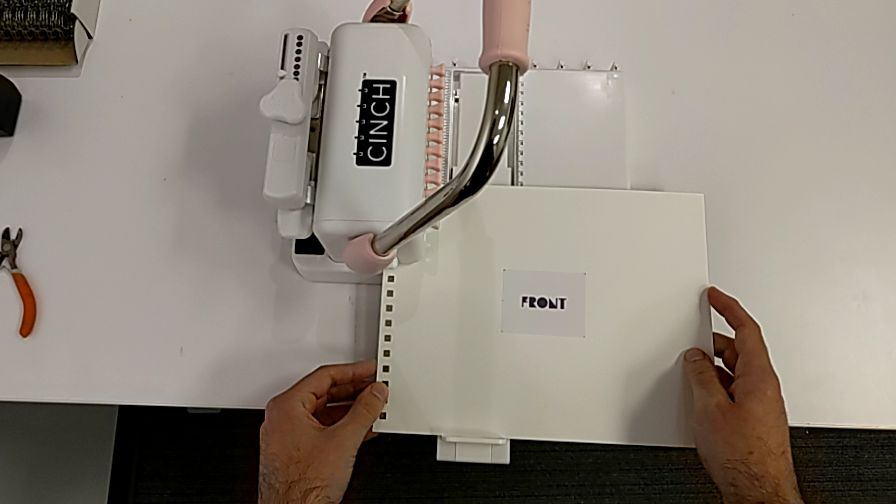}}
    \fbox{\includegraphics[width=4.4cm]{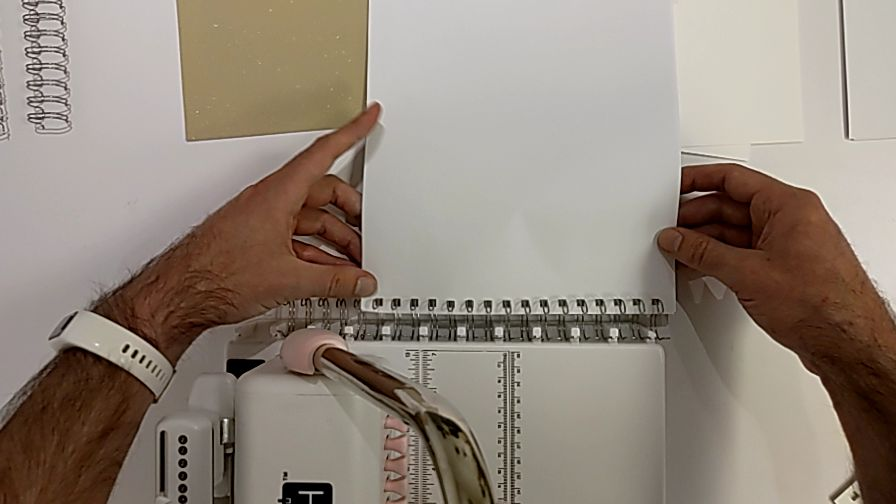}}
    \fbox{\includegraphics[width=4.4cm]{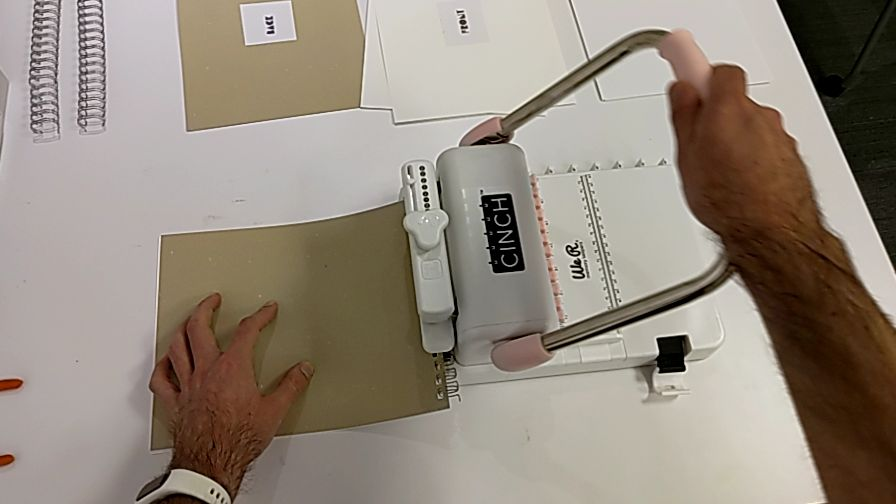}}
\end{figure}
\begin{enumerate}[label*=\arabic*.]
	\item Prepare the binding machine
	\begin{enumerate}[label*=\arabic*.]
		\item Unhook the storage hook from the handle bar. The handle bar should automatically lift once the hook is released. \textcolor{lightgray}{[Move]}
		\item Make sure all the pink pegs, from number 1 to 12 are pushed in. \textcolor{lightgray}{[Check]}
		\item Make sure the edge guide is pushed all the way into the machine. The edge guide is the vertical plastic edge connected to the ruler. \textcolor{lightgray}{[Check]}
	\end{enumerate}
	\item Punch the short edge of the covers
	\begin{enumerate}[label*=\arabic*.]
		\item Take a front and back cover and place them in the binding machine aligning with the edge guide. \textcolor{lightgray}{[Move]}
		\item Punch the short edge of the covers by pressing the handle bar down. Remove the covers from the machine. \textcolor{lightgray}{[Move]}
		\item Extend the edge guide all the way out. \textcolor{lightgray}{[Discrete Event]}
		\item Place the covers in the binding machine again, aligning them with the new edge guide placement. \textcolor{lightgray}{[Move]}
		\item To secure the covers in place, push the alignment peg down through the punched hole underneath it. The alignment peg is the white plastic knob underneath the handle. \textcolor{lightgray}{[Discrete Event]}
		\item Pull out pink peg number 5. \textcolor{lightgray}{[Discrete Event]}
		\item Press the handle bar again to punch the remaining holes. \textcolor{lightgray}{[Discrete Event]}
		\item Pull up the alignment peg to release the covers. Remove the covers from the machine. \textcolor{lightgray}{[Move]}
		\item Push the edge guide all the way back in. \textcolor{lightgray}{[Discrete Event]}
		\item Push the pink peg number 5 back in. \textcolor{lightgray}{[Discrete Event]}
	\end{enumerate}
	\item Punch the notebook pages
	\begin{enumerate}[label*=\arabic*.]
		\item Take 4 blank pages and place them in the binding machine aligning with the edge guide. \textcolor{lightgray}{[Move]}
		\item Punch the pages by pressing the handle bar down. Remove the pages from the machine. \textcolor{lightgray}{[Move]}
		\item Extend the edge guide all the way out. \textcolor{lightgray}{[Discrete Event]}
		\item Place the pages in the binding machine again, aligning them with the new edge guide placement. \textcolor{lightgray}{[Move]}
		\item To secure the pages in place, push the alignment peg down through the punched hole underneath it. \textcolor{lightgray}{[Discrete Event]}
		\item Pull out pink peg number 5. \textcolor{lightgray}{[Discrete Event]}
		\item Press the handle bar again to punch the remaining holes. \textcolor{lightgray}{[Discrete Event]}
		\item Pull up the alignment peg to release the pages. Remove the pages from the machine. \textcolor{lightgray}{[Move]}
		\item Push the edge guide all the way back in. \textcolor{lightgray}{[Discrete Event]}
		\item Push the pink peg number 5 back in. \textcolor{lightgray}{[Discrete Event]}
	\end{enumerate}
	\item Prepare the notebook for binding
	\begin{enumerate}[label*=\arabic*.]
		\item Place the binding wire on the plastic hooks on the edge of the machine. The binding wire has wide loops on one side and narrow loops on the other. Place the wide loops on the plastic hooks, with every other wide loop on a hook. Make sure the opening of the binding wire, or the empty part of the `C` shape it forms is oriented up. \textcolor{lightgray}{[Move]}
		\item Place the inside pages on the narrow loops of the binding wire. We will trim the excess wire later. \textcolor{lightgray}{[Move]}
		\item Thread the front cover onto the narrow loops of the binding wire, on top of the pages, with the outer face facing up. \textcolor{lightgray}{[Move]}
		\item Thread the back cover onto the narrow loops of the binding wire, this time with its outer face down, towards the front cover. \textcolor{lightgray}{[Move]}
		\item Remove the binding wire with the threaded notebook from the machine. \textcolor{lightgray}{[Move]}
		\item Use the wire cutters to trim the excess wire from the end of the notebook. \textcolor{lightgray}{[Discrete Event]}
	\end{enumerate}
	\item Bind the notebook
	\begin{enumerate}[label*=\arabic*.]
		\item Push down and turn the knob on top of the machine to adjust the binding size to 5/8 inches. \textcolor{lightgray}{[Discrete Event]}
		\item With the notebook still threaded, place the binding wire into the cinch press area of the machine, above the long black strip. Make sure to place the binding wire such that the open part, or the empty part of the `C` shape it forms, is flush against the vertical body of the machine. \textcolor{lightgray}{[Move]}
		\item Hold the binding wire in place by holding on to the paper, and do not put your fingers inside the press. Then, push the handle bar to cinch a portion of the wires. \textcolor{lightgray}{[Discrete Event]}
		\item Shift the notebook placement so that the un-cinched portion of the wires is now in the press. Make sure your fingers are out of the press, and push the handle bar again to cinch the remaining wires. \textcolor{lightgray}{[Discrete Event]}
		\item Remove the notebook from the press and fold back the back cover. \textcolor{lightgray}{[Move]}
	\end{enumerate}
\end{enumerate}
\newpage
\textbf{Task 6: Install the wheels on a skateboard}
\begin{figure}[h!]
    \centering
    \setlength{\fboxsep}{0pt}
    \fbox{\includegraphics[width=4.4cm]{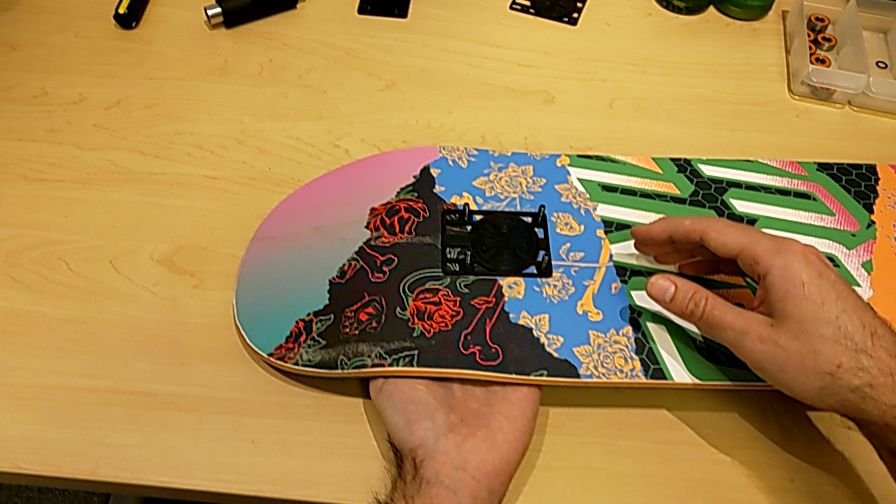}}
    \fbox{\includegraphics[width=4.4cm]{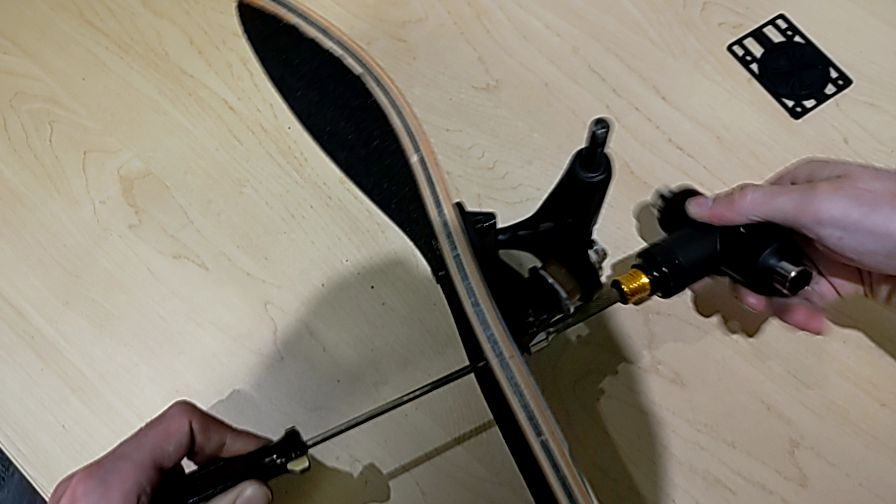}}
    \fbox{\includegraphics[width=4.4cm]{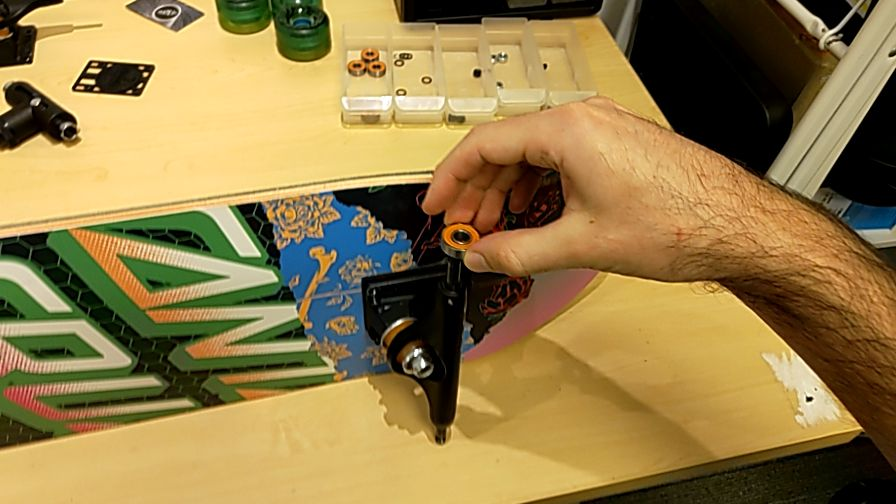}}
    \fbox{\includegraphics[width=4.4cm]{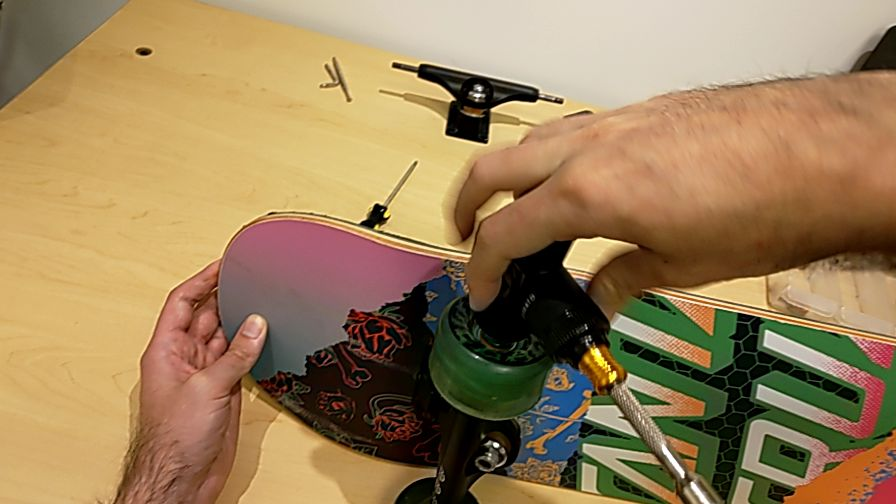}}
\end{figure}
\begin{enumerate}[label*=\arabic*.]
	\item Mount the truck
	\begin{enumerate}[label*=\arabic*.]
		\item Push four screws through the holes from the top of the board and then flip the board over so it's upside down on the table. \textcolor{lightgray}{[Move]}
		\item Position the riser onto the screws on the bottom of the deck with the slightly longer edge positioned outwards. \textcolor{lightgray}{[Discrete Event]}
		\item Position a truck over the riser and screws. The bushings should be facing inwards pointing to the middle of the skateboard. \textcolor{lightgray}{[Discrete Event]}
		\item Add a nut to each screw and tighten them with your fingers. \textcolor{lightgray}{[Invisible State]}
		\item Further tighten each screw. Use a screwdriver to hold each screw in place on the top of the deck while you use the skate tool to tighten the nut from the bottom. Do this for all four screws. \textcolor{lightgray}{[Invisible State]}
	\end{enumerate}
	\item Insert bearings into the wheels
	\begin{enumerate}[label*=\arabic*.]
		\item Tilt your board to the side and put one bearing on the axle of your truck. \textcolor{lightgray}{[Discrete Event]}
		\item Push your wheel hard into the bearing. You should feel the bearing click into place. \textcolor{lightgray}{[Invisible State]}
		\item Remove the wheel from the axle. \textcolor{lightgray}{[Move]}
		\item Install a second bearing into the other side of the wheel using the same procedure. Afterwards, remove the wheel from the axle. \textcolor{lightgray}{[Move]}
		\item Now tilt your board so you can use the other axle of your truck to install bearings on the second wheel. Put one bearing on that axle. \textcolor{lightgray}{[Discrete Event]}
		\item Push the second wheel hard into the bearing until it clicks into place, and then remove the wheel from the axle. \textcolor{lightgray}{[Move]}
		\item Install a second bearing into the other side of the wheel using the same procedure. Afterwards, remove the wheel from the axle. \textcolor{lightgray}{[Move]}
	\end{enumerate}
	\item Attach the first wheel to the truck
	\begin{enumerate}[label*=\arabic*.]
		\item Place a spacer on the axle. \textcolor{lightgray}{[Discrete Event]}
		\item Place a wheel on the same axle over the spacer. The graphic on the wheel should face outside. \textcolor{lightgray}{[Discrete Event]}
		\item Place another spacer on the axle, on the outside of the wheel. \textcolor{lightgray}{[Discrete Event]}
		\item Add a nut to the axle over the second spacer and use your fingers to tighten the nut. \textcolor{lightgray}{[Invisible State]}
		\item Use your skate tool to further tighten the nut, but make sure not to overtighten. At the end, the wheel should spin freely. \textcolor{lightgray}{[Invisible State]}
	\end{enumerate}
	\item Attach the second wheel to the truck
	\begin{enumerate}[label*=\arabic*.]
		\item Place a spacer on the second axle and place the second wheel on the axle over the spacer. \textcolor{lightgray}{[Discrete Event]}
		\item Place a second spacer on the axle, on the outside of the wheel, and then add a nut to the axle over the second spacer. Use your fingers to tighten the nut. \textcolor{lightgray}{[Invisible State]}
		\item Use your skate tool to further tighten the nut, but again, make sure not to overtighten. \textcolor{lightgray}{[Invisible State]}
	\end{enumerate}
	\item Make final adjustments
	\begin{enumerate}[label*=\arabic*.]
		\item Make sure the trucks are securely attached, and the nuts are tightened properly. \textcolor{lightgray}{[Check]}
		\item Adjust the kingpin nut if necessary to achieve the desired tightness or looseness for turning. \textcolor{lightgray}{[Check]}
		\item Spin each wheel to ensure they are spinning smoothly and there is no resistance or wobbling. \textcolor{lightgray}{[Invisible State]}
	\end{enumerate}
\end{enumerate}
\newpage
\textbf{Task 7: Make a Margarita mocktail}
\begin{figure}[h!]
    \centering
    \setlength{\fboxsep}{0pt}
    \fbox{\includegraphics[width=4.4cm]{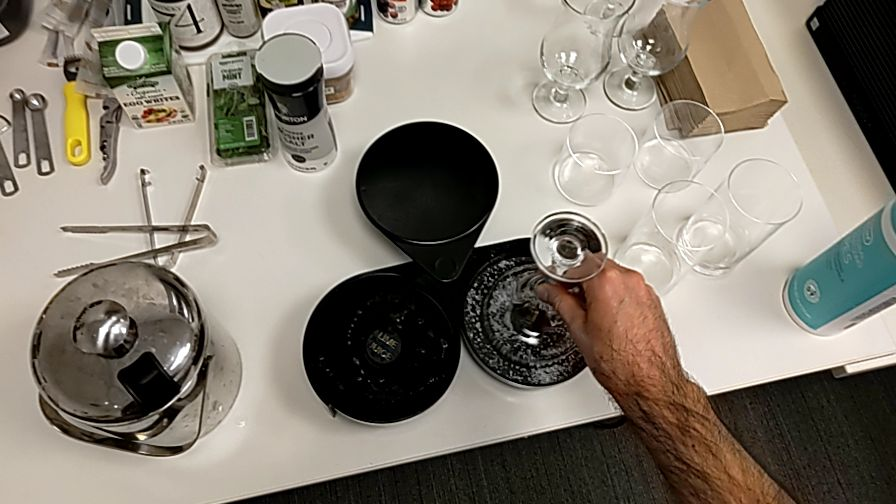}}
    \fbox{\includegraphics[width=4.4cm]{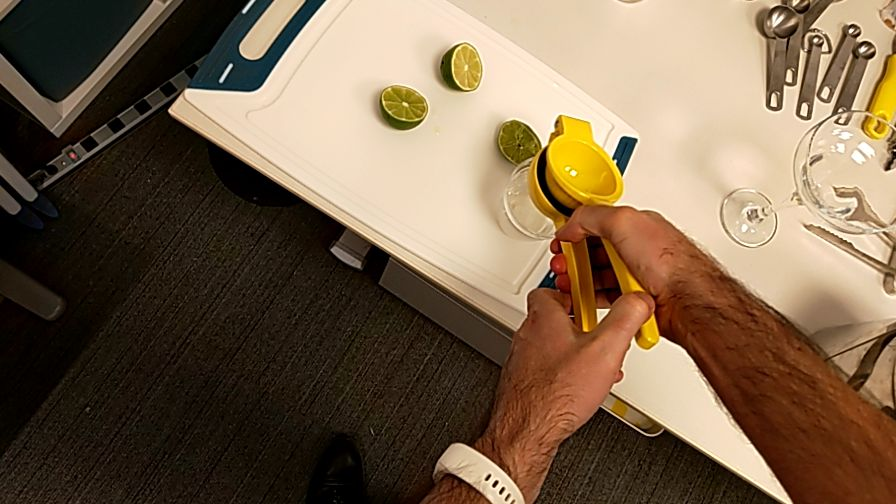}}
    \fbox{\includegraphics[width=4.4cm]{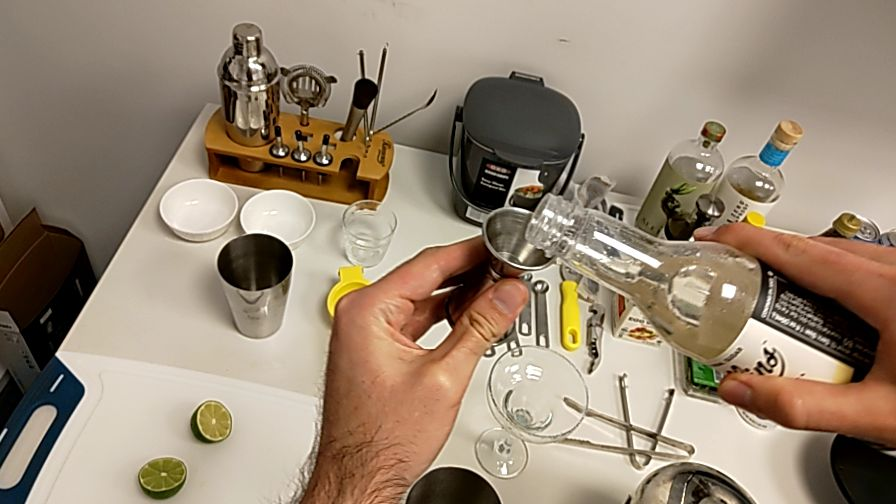}}
    \fbox{\includegraphics[width=4.4cm]{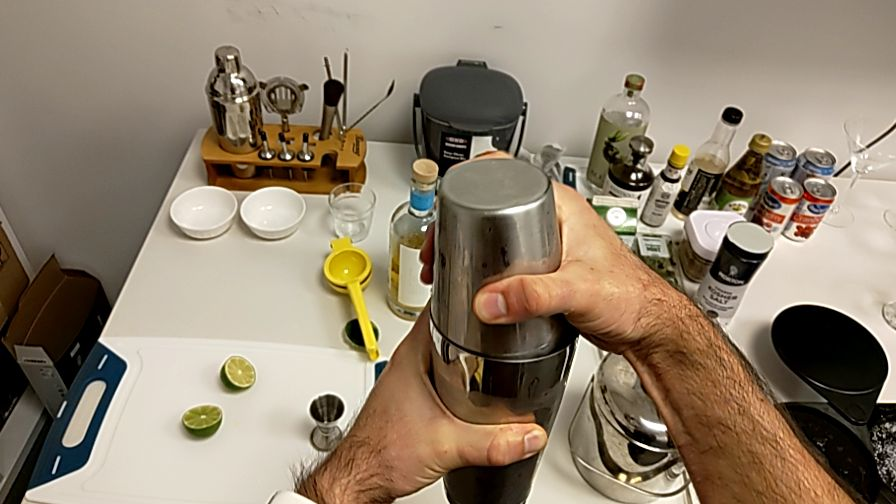}}
\end{figure}
\begin{enumerate}[label*=\arabic*.]
	\item Prepare the margarita glass with salt
	\begin{enumerate}[label*=\arabic*.]
		\item Squeeze half a lime onto the sponge of the lime section of the glass rimmer. \textcolor{lightgray}{[Add Amount]}
		\item Place salt in the salt section of the glass rimmer. Make sure it's evenly covering the surface. \textcolor{lightgray}{[Add Amount]}
		\item Take a margarita glass, hold it upside down, and tap and press the rim onto the sponge imbibed with lime juice. \textcolor{lightgray}{[Invisible State]}
		\item Tap the glass rim into the salt container. \textcolor{lightgray}{[Invisible State]}
	\end{enumerate}
	\item Prepare the ingredients
	\begin{enumerate}[label*=\arabic*.]
		\item Use the citrus squeezer to squeeze the second lime half into a small cup. \textcolor{lightgray}{[Add Amount]}
		\item Using the jigger, measure and pour 1 ounce of fresh lime juice into the larger tin of a Boston shaker. \textcolor{lightgray}{[Add Amount]}
		\item Pour in 0.5 ounces of simple syrup. \textcolor{lightgray}{[Add Amount]}
		\item Pour in 1.5 ounces of zero-proof tequila. \textcolor{lightgray}{[Add Amount]}
		\item Add 6 or 7 ice cubes in the tin. \textcolor{lightgray}{[Add Amount]}
	\end{enumerate}
	\item Shake and pour the cocktail
	\begin{enumerate}[label*=\arabic*.]
		\item Cover the large tin with the smaller tin, and push the top (smaller) tin down firmly to ensure a seal. \textcolor{lightgray}{[Invisible State]}
		\item Using both hands to catch the tins, make sure they stay pressed together and shake the drink 7 to ten seconds, until you feel that the metal has become quite cold. Make sure you hold them tight, as the shaking process might have them come undone and the drink might spill. \textcolor{lightgray}{[Invisible State]}
		\item Open the top tin. If it does not come out easily, you can tap it from the side, or try to rotate it up. \textcolor{lightgray}{[Move]}
		\item Place a hawthorne strainer over the opening of the large tin. \textcolor{lightgray}{[Move]}
		\item While holding the strainer over the top of the tin, tilt the tin and strain the drink into the prepared glass. \textcolor{lightgray}{[Add Amount]}
	\end{enumerate}
\end{enumerate}
\newpage
\textbf{Task 8: Make a whiskey sour mocktail}
\begin{figure}[h!]
    \centering
    \setlength{\fboxsep}{0pt}
    \fbox{\includegraphics[width=4.4cm]{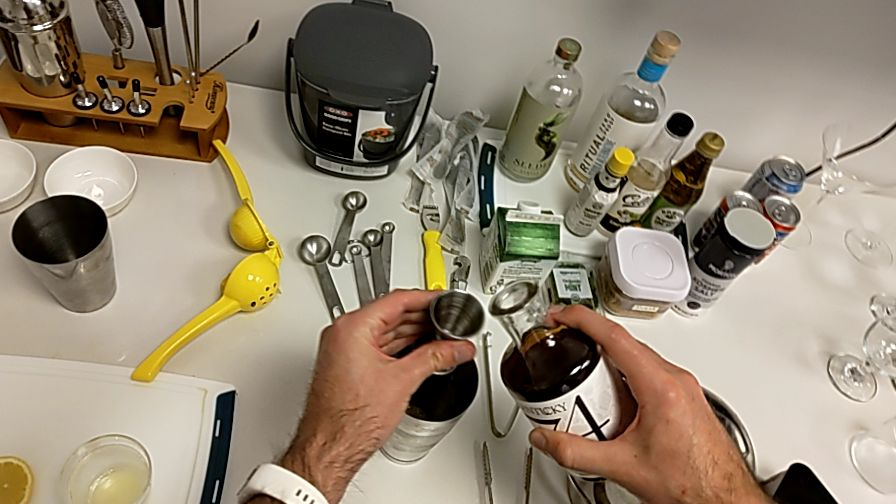}}
    \fbox{\includegraphics[width=4.4cm]{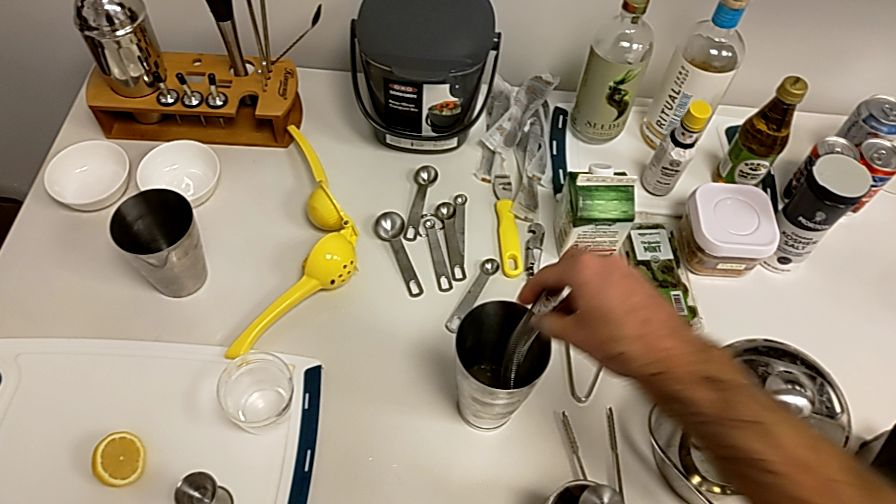}}
    \fbox{\includegraphics[width=4.4cm]{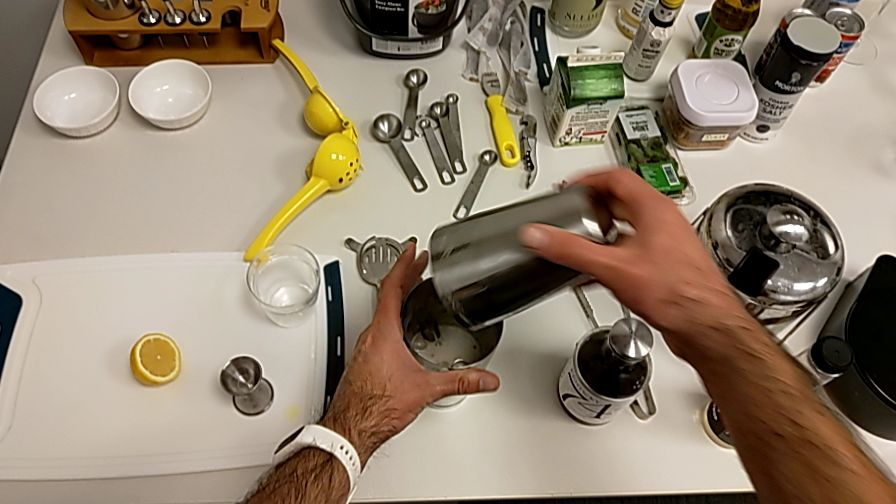}}
    \fbox{\includegraphics[width=4.4cm]{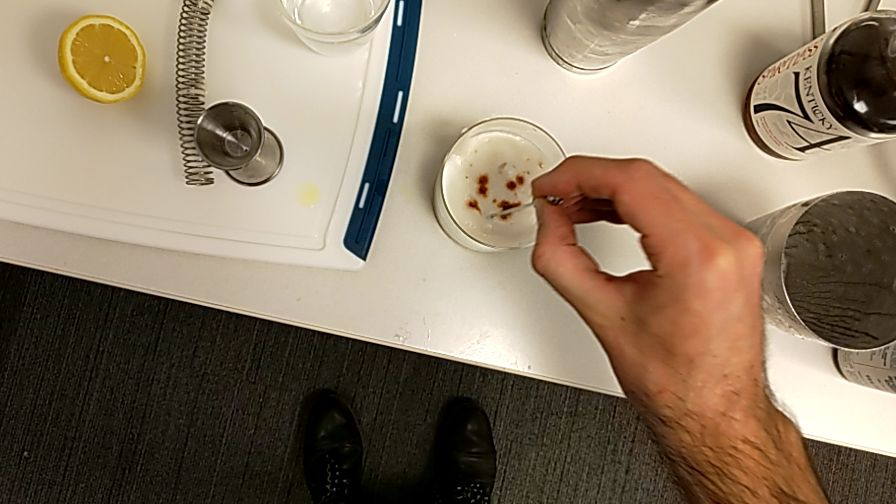}}
\end{figure}
\begin{enumerate}[label*=\arabic*.]
	\item Prepare the ingredients
	\begin{enumerate}[label*=\arabic*.]
		\item Using the jigger, measure and pour one ounce of egg white into the larger tin of a Boston shaker. \textcolor{lightgray}{[Add Amount]}
		\item Using the citrus squeezer, extract the juice from a lemon half into a cup. \textcolor{lightgray}{[Add Amount]}
		\item Pour 2 ounces of non-alcoholic Kentucky 74 bourbon into the tin with the egg white. \textcolor{lightgray}{[Add Amount]}
		\item Pour 3/4 ounces of the freshly squeezed lemon juice into the same tin. \textcolor{lightgray}{[Add Amount]}
		\item Pour half an ounce of simple syrup into the tin. \textcolor{lightgray}{[Add Amount]}
	\end{enumerate}
	\item Dry shake to froth the cocktail
	\begin{enumerate}[label*=\arabic*.]
		\item Take out the spring from the Hawthorne strainer and place it into the same tin. This will help beat the egg when we shake the cocktail. \textcolor{lightgray}{[Discrete Event]}
		\item Cover the large tin of the Boston shaker with the smaller tin, and push the top (smaller) tin down firmly to ensure a seal. \textcolor{lightgray}{[Invisible State]}
		\item Using both hands to catch the tins, make sure they stay pressed together and shake the drink for about 30 seconds. As you shake it the spring inside will froth the egg white. Make sure you hold them tight, as the shaking process might have them come undone and the drink might spill. \textcolor{lightgray}{[Invisible State]}
		\item Open the top tin. If it does not come out easily, you can tap it from the side, or try to rotate it up. The drink inside should look frothy. \textcolor{lightgray}{[Move]}
		\item Remove the spring from the tin and set it aside. \textcolor{lightgray}{[Move]}
	\end{enumerate}
	\item Shake the cocktail with ice to chill
	\begin{enumerate}[label*=\arabic*.]
		\item Add 6 or 7 ice cubes to the tin. \textcolor{lightgray}{[Add Amount]}
		\item Cover again the tin with the smaller tin and push the top down firmly to ensure a seal. \textcolor{lightgray}{[Invisible State]}
		\item Again, using both hands to make sure the tins stay pressed together, shake the drink for 7 to ten seconds, until the tins become cold to the touch. \textcolor{lightgray}{[Invisible State]}
		\item Open again the top tin. If it doesn’t open easily, tap it from the side, or try to rotate it up. \textcolor{lightgray}{[Move]}
	\end{enumerate}
	\item Strain the drink into the glass
	\begin{enumerate}[label*=\arabic*.]
		\item Place the Hawthorne strainer without the spring over the top of the tin. \textcolor{lightgray}{[Move]}
		\item While holding the Hawthorne strainer over the top of the tin, tilt and pour the drink into a rocks glass. Ideally, there should be some foam at the top. \textcolor{lightgray}{[Add Amount]}
	\end{enumerate}
	\item Garnish
	\begin{enumerate}[label*=\arabic*.]
		\item Place 3 or 4 drops of the Angostura bitters over the foam at the top of the mocktail. \textcolor{lightgray}{[Add Amount]}
		\item Using the metal toothpick, trace a line through the bitter drops on top of the foam. This will reshape the drops to look like little hearts. \textcolor{lightgray}{[Move]}
	\end{enumerate}
\end{enumerate}

\newpage

\section{Demographics Questionnaire}
\label{app:DemographicsQuestionnaire}
Below is the demographics questionnaire that each participant in the \textsc{SigmaCollab} data collection study completed at the end of the study.

\vspace{2cm}

1.	What is your age? (Circle one)

\setlength{\tabcolsep}{18pt}
\begin{tabular*}{\linewidth}{ccccccc}
18-25	& 26-35	& 36-45	& 46-55	& 56-65	& 66+ &   Prefer not to say
\end{tabular*}

\vspace{20pt}

2.	What is your gender? 

Man

Woman

Non-binary / gender diverse

Self-described: .....................

Prefer not to say

\vspace{20pt}

3.	What is your occupation?

\vspace{20pt}

4.	How familiar are you with mixed, augmented, and/or virtual reality technologies?

\begin{tabular*}{\linewidth}{ccccccccc}
Not familiar at all	&1&	2&	3&	4&	5&	6&	7&	Very familiar
\end{tabular*}

\vspace{20pt}
5.	How often do you use a VR or MR headset?

\begin{tabular*}{\linewidth}{ccccccccc}
Never	&1&	2&	3&	4&	5&	6&	7&	Often (daily)
\end{tabular*}

\vspace{10pt}
6.	Do you own a VR or MR headset?   $\square$ No  $\square$ Yes. What kind? .....................

\newpage

\end{document}